\documentclass[aps,prd,onecolumn,12pt,longbibliography]{revtex4-2}
\usepackage[table]{xcolor}
\usepackage{amsmath,amsthm}
\usepackage{amsfonts}
\usepackage{amssymb}
\usepackage{amsbsy,bm}

\theoremstyle{plain}
\newtheorem{theorem}{Theorem}

\theoremstyle{definition}

\theoremstyle{remark}

\newtheorem{example}[theorem]{Example}

\numberwithin{equation}{section}

\usepackage{exscale}
\usepackage[scaled=0.86]{helvet}
\usepackage{graphicx}

\usepackage[scr=euler,scrscaled=1.1,bb=txof,bbscaled=1.1,cal=boondox,calscaled=1.1]{mathalfa}
\renewcommand{\mathbf}{\bm}
\renewcommand{\mathit}[1]{\mathscr #1}
\renewcommand{\mathfrak}[1]{{\textsc{\upshape #1}}}
\renewcommand{\mathtt}[1]{\textcolor{red}{\texttt{\upshape#1}}}
\renewcommand{\mathrm}[1]{\scalebox{1.15}{\textbf{\textup{#1}}}}
\date{\today}
\usepackage{natbib}

\begin{document}

\title{Ternary public-key cryptosystem}

\author{\textbf{Steven Duplij}}
\email{douplii@uni-muenster.de, duplij@gmx.de, http://www.uni-muenster.de/IT.StepanDouplii}

\affiliation{
College of Information and Communication Engineering, Harbin Engineering University, 150001 Harbin, China}

\affiliation{Center for Information Technology,\\
University of M\"unster,
48149 M\"unster,
Germany}

\author{\textbf{Qiang Guo}}
\email{guoqiang292004@163.com, guoqiang@hrbeu.edu.cn}

\affiliation{
College of Information and Communication Engineering,
Harbin Engineering University,
150001 Harbin, China}

\author{\textbf{Na Fu}}
\email{nafu@hrbeu.edu.cn}

\affiliation{
College of Information and Communication Engineering,
Harbin Engineering University,
150001 Harbin, China}

\date{June 5, 2026}
\date{\today}
%\usepackage[usenames,dvipsnames]{xcolor}
%%\usepackage{color}
%\subjclass[2010]{16W25, 53B05, 53B15, 53D17, 81Q60, 81Q65, 81S99}
%\date{\textit{start} November 12, 2020, \textit{current}:
%\textit{of completion}
%{%\large
%\textbf{\today} at \currenttime}}

%\bigskip

\begin{abstract}

\noindent 
Public‑key cryptosystems eliminate the requirement for pre‑shared secret keys by enabling encryption with a publicly disclosed key and decryption with a corresponding private key. In this article we generalize the public‑key cryptosystems to ternary algebraic structures, with particular attention to ElGamal as a representative family.

We introduce the necessary algebraic background for nonderived ternary structures, including special elements, ternary group rings, and a matrix ternarization procedure that maps binary rings and group rings to antidiagonal symbolic matrices closed under ternary multiplication. Building on these foundations, we formulate a ternary analogue of the ElGamal three‑step protocol (key generation, ephemeral encryption, and decryption via querelements) and derive explicit ternary power and querelement formulas that enable correct decryption. Concrete instantiations and numerical examples over a ternary fraction field, a matrix‑ternarized finite group ring, and a finite \((6,3)\)-ring (field) validate the construction and illustrate admissible word‑length quantization and cycle behaviour of ternary powers.
The ternary framework highlights two practical advantages: richer algebraic structure (querelements replace binary inverses) that increases algebraic complexity for attackers, and higher information density (matrix ternarization transfers paired/plaintext vectors). Formal hardness assumptions, optimized parameter choices, and comprehensive security and performance analyses remain necessary future work.

\end{abstract}

\maketitle
\newpage
%\begin{small}
\tableofcontents
%\end{small}%
%\newpage

\section*{Introduction}

Public-key cryptography, introduced in the mid-1970s, removed the longstanding
requirement for a secure pre-shared channel to distribute secret keys by
separating encryption and decryption into distinct, mathematically linked
keys: a publicly disclosed encryption key and a privately held decryption key.
This separation enabled any party to encrypt a message for a recipient without
prior secret exchange and has become foundational to secure web protocols,
electronic payments, digital signatures, and private messaging. Diffie and
Hellman introduced the core idea of public-key exchange and motivated
subsequent asymmetric constructions \cite{dif/hel1976}, which remains the
conceptual basis for many discrete-log based schemes.

Early practical realizations of the paradigm include RSA, which bases security
on integer factorization and demonstrated the feasibility of public-key
encryption in practice \cite{riv/sha/adl1978}. Shortly thereafter, the ElGamal
cryptosystem~\cite{elg1985} offered an alternative construction rooted in the
discrete logarithm problem, introducing probabilistic encryption---a feature
that fundamentally distinguishes it from deterministic schemes. The
probabilistic nature of ElGamal---where identical plaintexts encrypt to
different ciphertexts with overwhelming probability---yields ciphertext
indistinguishability under appropriate group choices and underpins many modern
cryptographic constructions \cite{kat/lin2020}. ElGamal's algebraic pattern
also directly inspired practical standards such as the Digital Signature
Algorithm (DSA), which adapts ElGamal-style signatures for standardized use
\cite{nist2013}.

Motivated by the algebraic diversity of polyadic systems \cite{duplij2022},
this paper develops a ternary generalization of ElGamal-style public-key
encryption. We present the algebraic preliminaries for nonderived ternary
structures (special elements, ternary group rings, and a matrix ternarization
mapping), formulate an abstract ternary encryption/decryption protocol that
mirrors the three-step ElGamal flow (key generation, ephemeral encryption, and
decryption), and validate the construction with concrete examples. The ternary
setting replaces binary powers and inverses with $\ell$-ternary powers and
querelements (the ternary analogue of inverses), drawing on the formalism of
$\left(  m,n\right)  $-rings and fields \cite{duplij2022}. To obtain practical
instantiations we employ a matrix ternarization that maps binary rings and
group rings to antidiagonal symbolic matrices closed under ternary
multiplication, this construction is informed by prior work on group rings
\cite{dup2026h} and matrix realizations of $n$-ary algebras \cite{dup2022a}.

The remainder of the paper details the necessary ternary algebra, the matrix
ternarization procedure that produces nonderived ternary unit groups, the
ternary analogue of the ElGamal protocol with explicit formulas for encryption
and decryption, and several numerical examples that illustrate correctness and
structural properties relevant to security and implementation.

\section{Ternary structures}

We present the notation and general properties of ternary structures (see
\cite{duplij2022} for details). Let $S^{\times3}$ be the $3$-fold Cartesian
product of a non-empty set $S$. Elements $(x_{1},x_{2},x_{3})\in S^{\times3}$
are called $3$-tuples $(\mathfrak{x})$; an $3$-tuple of identical elements is
denoted $(x^{3})$. A ternary operation is a mapping $\mu^{\left[  3\right]
}:S^{\times3}\rightarrow S$, written $\mu^{\left[  3\right]  }[\mathfrak{x}]$.
In general, a ternary structure $\langle S\mid{\mu}_{i}^{\left[  3\right]
}\rangle$ consists of a set $S$ closed under a family of the operations ${\mu
}_{i}^{\left[  3\right]  }$.

\subsection{Special elements in ternary groups and rings\label{subsec-special}%
}

The fundamental one-operation structure is the ternary magma $\mathcal{M}%
^{\left[  3\right]  }=\langle S\mid\mu^{\left[  3\right]  }\rangle$. Adding
axioms gives group-like structures: a ternary associative magma is a ternary
semigroup $\mathcal{S}^{\left[  3\right]  }=\langle S\mid\mu^{\left[
3\right]  }\mid assoc\rangle$. Ternary associativity means%
\begin{equation}
\mu^{\left[  3\right]  }\left[  \mu^{\left[  3\right]  }\left[  x_{1}%
,x_{2},x_{3}\right]  ,x_{4},x_{5}\right]  =\mu^{\left[  3\right]  }\left[
x_{1},\mu^{\left[  3\right]  }[x_{2},x_{3},x_{4}],x_{5}\right]  =\mu^{\left[
3\right]  }\left[  x_{1},x_{2},\mu^{\left[  3\right]  }\left[  x_{3}%
,x_{4},x_{5}\right]  \right]
\end{equation}
and so it is invariant under placing the inner multiplication in any of the
$2$ positions (giving $2$ relations), allowing omission of parentheses. The
iterated product is
\begin{equation}
\mu^{\left[  3\right]  \circ\ell_{\mu}}[\mathfrak{x}]=\overset{\ell_{\mu}%
}{\overbrace{\mu^{\left[  3\right]  }[\mu^{\left[  3\right]  }[\ldots
\mu^{\left[  n\right]  }[\mathfrak{x}]]]}},\quad\mathfrak{x}\in S^{2\ell_{\mu
}+1}, \label{ml}%
\end{equation}
where $\ell_{\mu}$ is the number of composed ternary multiplications
$\mu^{\left[  3\right]  }$. If $n=2$, then the iterated product (\ref{ml}) of
arity $\ell_{\mu}+1$ is called derived.

A key distinction from the binary case ($n=2$) is that the word length $w$ in
a composition of $n$-ary multiplications is quantized and depends of arity:
multiplication is possible for
\begin{equation}
w^{\text{admiss}}(n,\ell_{\mu})=\left(  n-1\right)  \ell_{\mu}+1, \label{wln}%
\end{equation}
and for the ternary case%
\begin{equation}
w^{\text{admiss}}(\ell_{\mu})=2\ell_{\mu}+1 \label{wl}%
\end{equation}
elements. This leads to two types of ternary operations: derived (iterated
from binary/lower-arity operations) and nonderived; the latter are more interesting.

For $x\in S$, the $\ell_{\mu}$-ternary power is
\begin{equation}
x^{\langle\ell_{\mu}\rangle}=\mu^{\left[  3\right]  \circ\ell_{\mu}}%
[x^{2\ell_{\mu}+1}], \label{xl}%
\end{equation}
which for $n=2$ gives $x^{\langle\ell_{\mu}\rangle}=x^{\ell_{\mu}+1}$,
differing by one from the usual power. A ternary idempotent $x_{id}$ satisfies
$x_{id}^{\langle\ell_{\mu}\rangle}=x_{id}$. A ternary zero $z$ is uniquely
defined by the $3$ conditions%
\begin{equation}
{\mu}^{\left[  3\right]  }[z,x_{1},x_{2}]={\mu}^{\left[  3\right]  }%
[x_{1},z,x_{2}]={\mu}^{\left[  3\right]  }[x_{1},x_{2},z]=z,\ \ \ \ \ x_{1}%
,x_{2},z\in S. \label{z}%
\end{equation}
A ternary nilpotent $x_{nil}$ obeys
\begin{equation}
x_{nil}^{\langle\ell_{\mu}\rangle}=z. \label{xz}%
\end{equation}
A neutral $2$-tuple $\mathfrak{e}$ satisfies ${\mu}^{\left[  3\right]
}[x,\mathfrak{e}]=x$ (typically non-unique). If all components of
$\mathfrak{e}$ are identical, $\mathfrak{e}=e^{2}$, then the identity is
defined by%
\begin{equation}
\mu^{\left[  3\right]  }[x,e^{2}]=\mu^{\left[  3\right]  }[x,e,x]=\mu^{\left[
3\right]  }[e^{2},x]=x \label{e}%
\end{equation}
where $e$ may appear in any 2 of the $3$ positions. Obviously, that both $z$
and $e$ are idempotents. Some ternary structures lack idempotents, zero, or
identity, or have multiple identities.

For ternary structures, invertibility is not tied to the identity but to the
querelement $\bar{x}=\bar{x}(x)$ (or $\bar{x}=Q\left(  x\right)  $), defined
by the $3$ relations%
\begin{equation}
\mu^{\left[  3\right]  }[x^{2},\bar{x}]=\mu^{\left[  3\right]  }[x,\bar
{x},x]=\mu^{\left[  3\right]  }[x^{2},\bar{x}]=x\label{q}%
\end{equation}
(holding for $\bar{x}$ in each of the $3$ positions). Such $x$ is ternary
invertible. If every element of an ternary semigroup $\mathcal{S}_{3}$ is
ternary invertible, then $\mathcal{S}_{3}$ is an ternary group $\mathsf{G}%
^{\left[  3\right]  }=\langle G\mid\mu^{\left[  3\right]  }\mid assoc\rangle$,
where $G$ is the underlying set. The identity and neutral element (necessary
in binary groups) are not required for a ternary group $\mathsf{G}^{\left[
3\right]  }$.

Structures with two operations are ring-like. A generic $(m,n)$-ring,
$\mathcal{R}^{\left[  m,n\right]  }=\langle R\mid\nu^{\left[  m\right]  }%
,\mu^{\left[  n\right]  }\rangle$ consists of the underlying set $R$ with an
$m$-ary addition $\nu^{\left[  m\right]  }:R^{m}\rightarrow R$ and an $n$-ary
multiplication $\mu^{\left[  n\right]  }:R^{n}\rightarrow R$, such that
$\langle R\mid\nu^{\left[  m\right]  }\mid assoc\mid comm\rangle$ is an
$m$-ary commutative group and $\langle R\mid\mu^{\left[  n\right]  }\mid
assoc\rangle$ is an $n$-ary semigroup, linked by $n$-ary distributivity
relations (for all positions of the sum in the \cite{lee/but}. The
$(m,n)$-ring $\mathcal{R}^{\left[  m,n\right]  }$ is ternary derived, if both
operations are composed from ternary operations (\ref{ml}). For further
details, see \cite{duplij2022}.

\subsection{Ternary group rings}

The binary group rings were studied in \cite{bovdi,passman,sehgal}, while
$\left(  m,n\right)  $-group rings were introduced and investigated in
\cite{dup2026h}. We here remind in brief definitions to fix notations.

Let us consider $\left(  m_{r},n_{r}\right)  $-ring $\mathcal{R}^{\left[
m_{r},n_{r}\right]  }$ and $n_{g}$-ary group $\mathsf{G}^{\left[
n_{g}\right]  }$, then the general $\left(  \mathbf{m}_{r},\mathbf{n}%
_{r}\right)  $-group ring $\mathrm{R}^{\left[  \mathbf{m}_{r},\mathbf{n}%
_{r}\right]  }$ is denoted by%
\begin{equation}
\mathrm{R}^{\left[  \mathbf{m}_{r},\mathbf{n}_{r}\right]  }=\mathcal{R}%
^{\left[  n_{r},m_{r}\right]  }\left[  \mathsf{G}^{\left[  \mathsf{n}%
_{g}\right]  }\right]  ,\label{rm}%
\end{equation}
and is the 3 set and 6 operation algebra-like structure%
\begin{equation}
\mathrm{R}^{\left[  \mathbf{m}_{r},\mathbf{n}_{r}\right]  }=\left\langle
\mathfrak{R},R,G\mid\mathbf{\nu}_{\mathfrak{R}}^{\left[  \mathbf{m}%
_{r}\right]  },\mathbf{\mu}_{\mathfrak{R}}^{\left[  \mathbf{n}_{r}\right]
},\mathbf{\rho}_{\mathfrak{R}}^{\left[  \mathbf{k}_{\rho}\right]  }\mid\nu
_{R}^{\left[  m_{r}\right]  },\mu_{R}^{\left[  n_{r}\right]  }\mid
\mu_{\mathsf{G}}^{\left[  \mathsf{n}_{g}\right]  }\right\rangle ,\label{rmn}%
\end{equation}
where $\mu_{\mathsf{G}}^{\left[  \mathsf{n}_{g}\right]  }$ is $\mathsf{n}_{g}%
$-ary multiplication of the initial group $\mathsf{G}^{\left[  n_{g}\right]
}$ (with the underlying set $G$), $\nu_{R}^{\left[  m_{r}\right]  }$ and
$\mu_{R}^{\left[  n_{r}\right]  }$ are $m_{r}$-ary addition and $n_{r}$-ary
multiplication of the initial $\left(  m_{r},n_{r}\right)  $-ring
$\mathcal{R}^{\left[  n_{r},m_{r}\right]  }$ (with the underlying set $R$),
$\mathbf{\nu}_{\mathfrak{R}}^{\left[  \mathbf{m}_{r}\right]  }$, $\mathbf{\mu
}_{\mathfrak{R}}^{\left[  \mathbf{n}_{r}\right]  }$ and $\mathbf{\rho
}_{\mathfrak{R}}^{\left[  \mathbf{k}_{\rho}\right]  }$ are $\mathbf{\nu
}_{\mathfrak{R}}^{\left[  \mathbf{m}_{r}\right]  }$-ary addition,
$\mathbf{\mu}_{\mathfrak{R}}^{\left[  \mathbf{n}_{r}\right]  }$-ary
multiplication and $\mathbf{k}_{\rho}$-ary multiplication on scalar of the
$\left(  \mathbf{m}_{r},\mathbf{n}_{r}\right)  $-group ring $\mathrm{R}%
^{\left[  \mathbf{m}_{r},\mathbf{n}_{r}\right]  }$ (with the underlying set
$\mathfrak{R}$). The $\alpha$th element $\mathbf{r}\left(  \alpha\right)  $ of
the group ring (or the element of its underlying set $\mathfrak{R}$) can be
written as the formal sum (analog of the standard binary case $\sum
_{\mathsf{g}\in\mathsf{G}}r_{\mathsf{g}}\bullet\mathsf{g},\ \ \ r_{\mathsf{g}%
}\in R,\ \ \mathsf{g}\in G$)%
\begin{equation}
\mathrm{r}\left(  \alpha\right)  =\mathrm{r}\left(  \overrightarrow
{r}_{\overrightarrow{\mathsf{g}}}\left(  \alpha\right)  ,\overrightarrow
{\mathsf{g}}\right)  =\underset{i}{\mathbf{\Sigma}}\ r_{\mathsf{g}_{i}}\left(
\alpha\right)  \bullet\mathsf{g}_{i},\ \ \ \ r_{\mathsf{g}_{i}}\left(
\alpha\right)  \in R,\ \ \mathsf{g}_{i}\in G,\ \ \mathrm{r}\left(
\alpha\right)  \in\mathfrak{R},\label{ra}%
\end{equation}
where $r_{\mathsf{g}_{i}}\left(  \alpha\right)  $ are ring elements and
$\mathsf{g}_{i}$ are group elements. For conciseness, we denote the ordered
families $\overrightarrow{r}_{\overrightarrow{\mathsf{g}}}\left(
\alpha\right)  =\left(  r_{\mathsf{g}_{1}}\left(  \alpha\right)
,r_{\mathsf{g}_{2}}\left(  \alpha\right)  ,\ldots\right)  $, $\overrightarrow
{\mathsf{g}}=\left(  \mathsf{g}_{1},\mathsf{g}_{2},\ldots\right)  $ as
\textquotedblleft formal vectors\textquotedblright. In case the $\left(
\mathbf{m}_{r},\mathbf{n}_{r}\right)  $-group ring (\ref{rm}) is finite,
$\alpha=1,2\ldots\left\vert \mathfrak{R}\right\vert $.

In the $\left(  \mathbf{m}_{r},\mathbf{n}_{r}\right)  $-group ring
$\mathrm{R}^{\left[  \mathbf{m}_{r},\mathbf{n}_{r}\right]  }$, the
$\mathbf{m}_{r}$-ary addition $\mathbf{\nu}^{\left[  \mathbf{m}_{r}\right]  }$
is defined by analogy with the binary case, $\left(  \sum_{\mathsf{g}%
\in\mathsf{G}}r_{\mathsf{g}}\bullet\mathsf{g}\right)  +\left(  \sum
_{\mathsf{g}\in\mathsf{G}}r_{\mathsf{g}}^{\prime}\bullet\mathsf{g}\right)
=\left(  \sum_{\mathsf{g}\in\mathsf{G}}\left(  r_{\mathsf{g}}+r_{\mathsf{g}%
}^{\prime}\right)  \bullet\mathsf{g}\right)  $, $r_{\mathsf{g}}\in R$,
$\mathsf{g},\mathsf{h}\in G$, left-\textquotedblleft
componentwise\textquotedblright%
\begin{align}
\mathbf{\nu}_{\mathfrak{R}}^{\left[  \mathbf{m}_{r}\right]  }\left[
\mathrm{r}\left(  \overrightarrow{r}_{\overrightarrow{\mathsf{g}}}\left(
\alpha_{1}\right)  ,\overrightarrow{\mathsf{g}}\right)  ,\ldots,\mathrm{r}%
\left(  \overrightarrow{r}_{\mathsf{\vec{g}}}\left(  \alpha_{\mathbf{m}_{r}%
}\right)  ,\overrightarrow{\mathsf{g}}\right)  \right]   &  =\underset
{i}{\mathbf{\Sigma}}\ \nu_{R}^{\left[  m_{r}\right]  }\left[  r_{\mathsf{g}%
_{i}}\left(  \alpha_{1}\right)  ,\ldots,r_{\mathsf{g}_{i}}\left(
\alpha_{m_{r}}\right)  \right]  \bullet\mathsf{g}_{i},\label{mm}\\
r_{\mathsf{g}_{i}}\left(  \alpha_{1,\ldots,\mathbf{m}_{r}}\right)   &  \in
R,\ \ \mathsf{g}_{i}\in G,\ \ \ \mathrm{r}\left(  \overrightarrow
{r}_{\overrightarrow{\mathsf{g}}}\left(  \alpha_{1,\ldots,\mathbf{m}_{r}%
}\right)  ,\overrightarrow{\mathsf{g}}\right)  \in\mathfrak{R}.\nonumber
\end{align}

It follows from (\ref{mm}), that the arity of addition in the $\left(
\mathbf{m}_{r},\mathbf{n}_{r}\right)  $-group ring $\mathrm{R}^{\left[
\mathbf{m}_{r},\mathbf{n}_{r}\right]  }$ coincides with the arity of addition
in the initial $\left(  m_{r},n_{r}\right)  $-ring $\mathcal{R}^{\left[
n_{r},m_{r}\right]  }$ that is%
\begin{equation}
\mathbf{m}_{r}=m_{r}.\label{mrm}%
\end{equation}

If the $m_{r}$-ary addition in the initial $\left(  m_{r},n_{r}\right)  $-ring
$\mathcal{R}^{\left[  m_{r},n_{r}\right]  }$ is of higher ternary power
$\left(  \nu_{R}^{\left[  3\right]  }\right)  ^{\circ\ell_{m}}$, see
(\ref{ml}), then instead of (\ref{mrm}), we have the \textquotedblleft
quantization\textquotedblright%
\begin{equation}
\mathbf{m}_{r}=2\ell_{m}+1, \label{mlm}%
\end{equation}
which directly follows from the general form of admissible length of words
(\ref{wl}).

We define the multiplication in the $\left(  \mathbf{m}_{r},\mathbf{n}%
_{r}\right)  $-group ring $\mathrm{R}^{\left[  \mathbf{m}_{r},\mathbf{n}%
_{r}\right]  }$ (\ref{rm}) similarly to the binary case, $\left(
\sum_{\mathsf{g}\in G}r_{\mathsf{g}}\bullet\mathsf{g}\right)  \left(
\sum_{\mathsf{h}\in G}r_{\mathsf{h}}^{\prime}\bullet\mathsf{h}\right)
=\left(  \sum_{\mathsf{g}\in G}\sum_{\mathsf{h}\in G}\left(  r_{\mathsf{g}%
}r_{\mathsf{h}}^{\prime}\right)  \bullet\left(  \mathsf{gh}\right)  \right)
$, $r_{\mathsf{g}}\in R$, $\mathsf{g},\mathsf{h}\in G$, by the both-sides
\textquotedblleft componentwise\textquotedblright\ convolution-like operation%
\begin{align}
&  \mathbf{\mu}_{\mathfrak{R}}^{\left[  \mathbf{n}_{r}\right]  }\left[
\mathrm{r}\left(  \overrightarrow{r}_{\overrightarrow{\mathsf{g}}}\left(
\alpha_{1}\right)  ,\overrightarrow{\mathsf{g}}\right)  ,\ldots,\mathrm{r}%
\left(  \overrightarrow{r}_{\overrightarrow{\mathsf{g}}}\left(  \alpha
_{\mathbf{n}_{r}}\right)  ,\overrightarrow{\mathsf{g}}\right)  \right]
\nonumber\\
&  =\underset{i_{1}}{\mathbf{\Sigma}}\ \ldots\ \underset{i_{n_{r}}%
}{\mathbf{\Sigma\ }}\underset{j_{1}}{\mathbf{\Sigma}}\ \ldots\ \underset
{j_{\mathsf{n}_{g}}}{\mathbf{\Sigma}}\mu_{R}^{\left[  n_{r}\right]  }\left[
r_{\mathsf{g}_{i_{1}}}\left(  \alpha_{1}\right)  ,\ldots,r_{\mathsf{g}%
_{i_{n_{r}}}}\left(  \alpha_{n_{r}}\right)  \right]  \bullet\mu_{\mathsf{G}%
}^{\left[  \mathsf{n}_{g}\right]  }\left[  \mathsf{g}_{j_{1}},\ldots
,\mathsf{g}_{j_{\mathsf{n}_{g}}}\right]  ,\label{mng}\\
&  r_{\mathsf{g}_{i}}\left(  \alpha_{j}\right)  \in R,\ \ \mathsf{g}_{i}\in
G,\ \ \ \mathrm{r}\left(  \overrightarrow{r}_{\overrightarrow{\mathsf{g}}%
}\left(  \alpha_{j}\right)  ,\overrightarrow{\mathsf{g}}\right)
\in\mathfrak{R},\nonumber
\end{align}
if ternary power of all operations is $1$, such that the arities of all
multiplications defined in (\ref{mng}) coincide%
\begin{equation}
\mathbf{n}_{r}=n_{r}=\mathsf{n}_{g}.\label{nnn}%
\end{equation}

In case, the ternary powers of multiplications in the initial ternary ring
$\mu_{R}^{\left[  n_{r}\right]  }$ and the $\mathsf{n}_{g}$-ary group
$\mu_{\mathsf{G}}^{\left[  \mathsf{n}_{g}\right]  }$, denoted by $\ell_{n}$
and $\ell_{g}$, correspondingly, differ from $1$, (see (\ref{xl})), then we
have for the arity of multiplication $\mathbf{n}_{r}$ in the $\left(
\mathbf{m}_{r},\mathbf{n}_{r}\right)  $-group ring $\mathrm{R}^{\left[
\mathbf{m}_{r},\mathbf{n}_{r}\right]  }$, instead of (\ref{nnn}),%
\begin{equation}
\mathbf{n}_{r}=2\ell_{n}+1=2\ell_{g}+1, \label{nln}%
\end{equation}
which follows from the construction (\ref{mng}) and the \textquotedblleft
quantization\textquotedblright\ condition (\ref{wl}).

It is well known (see, e.g., \cite{duplij2022}) that the invertibility
properties of polyadic structures are governed not by neutral elements
(\ref{e}), but by querelements (\ref{q}). Let us consider the subset
$\emph{U}_{R}\subset R$ having the querelement $\bar{r}$, so such elements
form a multiplicative $\mathfrak{n}_{r}$-ary unit group $\mathit{U}^{\left[
n_{r}\right]  }\left[  \mathcal{R}^{\left[  m_{r},n_{r}\right]  }\right]  $ ),
while the $n_{g}$-ary group $\mathsf{G}^{\left[  \mathsf{n}_{g}\right]  }$
having (by its definition) the querelement $\mathsf{\bar{g}}$ (\ref{q}) for
each $\mathsf{g}\in\mathsf{G}$. Then in the group ring $\mathrm{R}^{\left[
\mathbf{m}_{r},\mathbf{n}_{r}\right]  }$ the querelement $\overline
{\mathrm{r}}\left(  \alpha\right)  $ for some elements is defined by%
\begin{equation}
\overline{\mathrm{r}}\left(  \alpha\right)  =\underset{i}{\mathbf{\Sigma}%
}\ \bar{r}_{\mathsf{g}_{i}}\left(  \alpha\right)  \bullet\mathsf{\bar{g}}%
_{i},\ \ \ \ \ \ \ \bar{r}\in\emph{U},\ \ \mathsf{\bar{g}}_{i},\mathsf{g}%
_{i}\in G,\ \ \overline{\mathrm{r}}\in\mathfrak{R},\label{rq}%
\end{equation}
and we call them multiplicatively invertible elements (sometimes called group
units), denote their subset $\emph{U}_{\mathfrak{R}}\in\mathfrak{R}$, which
forms the $\mathbf{n}_{r}$-ary unit group $\mathit{U}^{\left[  \mathbf{n}%
_{r}\right]  }\left(  \mathrm{R}^{\left[  \mathbf{m}_{r},\mathbf{n}%
_{r}\right]  }\right)  $ of the $\left(  \mathbf{m}_{r},\mathbf{n}_{r}\right)
$-group ring $\mathrm{R}^{\left[  \mathbf{m}_{r},\mathbf{n}_{r}\right]  }$.

The concept of ternary unit group is extremely important for cryptosystems,
because the decoding procedure can be made only by group units which
correspond to the binary multiplicatively invertible elements of (ternary)
rings and group rings.

\subsection{Matrix ternarization}

The structure of binary rings is well known (see, e.g.,
\cite{herstein,lam1991,haz/gub}): a simple ring (having no nontrivial
two-sided ideals) is isomorphic to the full matrix ring $Mat^{full}\left(
\mathcal{D}\right)  $ over a division ring $\mathcal{D}$ (each nonzero element
has inverse, or field, if $\mathcal{D}$ is commutative). For the simple
$\left(  2,n\right)  $-ring with binary addition and $n$-ary multiplication
$\mu^{\left[  n\right]  }$ (as ordinary product of matrices) the structure of
$\mathcal{R}_{simple}^{\left[  2,n\right]  }$ was obtained in \cite{nik84a}.
In the ternary case it is isomorphic to the antidiagonal $2\times2$ matrices
with blocks $Mat^{full}\left(  \mathcal{D}\right)  $ of the form%
\begin{equation}
\mathbf{M}_{adiag}^{\left[  3\right]  }\left(  \mathcal{D}\right)  =\left(
\begin{array}
[c]{cc}%
0 & Mat^{full}\left(  \mathcal{D}\right) \\
Mat^{full}\left(  \mathcal{D}\right)  & 0
\end{array}
\right)  . \label{mb}%
\end{equation}
The matrices of the form (\ref{mb}) are not closed with respect to the binary
product, but closed with respect to exactly ternary multiplication and binary
addition, and therefore we call them ternary matrices. Note that the ternary
multiplication $\mu^{\left[  3\right]  }$ is nonderived, because it cannot be
directly iterated (\ref{ml}) (without using additional homomorphisms, as in
\cite{glu1,glu64,hos}) from binary product which is not defined (not closed)
for the ternary matrices $\mathbf{M}_{adiag}^{\left[  3\right]  }\left(
\mathcal{D}\right)  $.

Moreover, a finite-dimensional simple associative ternary algebra
$\mathcal{A}^{\left(  3\right)  }$ over an algebraically closed field
$\mathcal{F}$ is isomorphic to the ternary matrix (\ref{mb}) $\mathbf{M}%
_{adiag}^{\left[  3\right]  }\left(  \mathcal{F}\right)  $ \cite{car4}.

Thus, going in the opposite way, we propose the matrix ternarization procedure
for binary algebraic structures, firstly introduced in \cite{dup2022a} for
general case, which can be considered as immediate and nonderived ternary
generalization of the known ordinary (binary) cryptosystems.

Let $B$ be a binary structure, for instance, semigroup, group, ring, group
ring. We defined the matrix ternarization as the mapping of the binary
structure $B$ to the symbolic matrix%
\begin{equation}
\mathbf{\Phi}_{\mathbf{M}}:B\rightarrow\widehat{\mathbf{M}}^{\left[  3\right]
}\left(  B\right)  , \label{f}%
\end{equation}
where%
\begin{equation}
\widehat{\mathbf{M}}^{\left[  3\right]  }\left(  B\right)  \equiv
\widehat{\mathbf{M}}_{adiag}^{\left[  3\right]  }\left(  B\right)  =\left(
\begin{array}
[c]{cc}%
0 & B\\
B & 0
\end{array}
\right)  . \label{mbb}%
\end{equation}
In this way, the multiplicative arity $n_{B}$ of $B$ can be also $3$, because
matrices of the form (\ref{mb}) are closed with respect to ternary
multiplication. The additive properties of $B$ are transferred to the symbolic
matrix addition without restrictions because of linearity.

The symbolic matrix construction of the ternary group ring from a binary group
ring $\mathcal{R}\left[  \mathsf{G}\right]  $ is more nontrivial: we should
use two symbolic matrices, one for the ring $\mathcal{R}$ and another for the
group $\mathsf{G}$. If their sizes coincide, the resulting matrix will have
the same size, and all the arities coincide. In this case $\mathbf{\Phi
}_{\mathbf{M}}$ (\ref{f}) is a homomorphism-like mapping, because using
(\ref{nln})
\begin{equation}
\widehat{\mathbf{M}}^{\left[  \mathbf{3}\right]  }\left(  \mathcal{R}\left[
\mathsf{G}\right]  \right)  =\widehat{\mathbf{M}}^{\left[  3\right]  }\left(
\mathcal{R}\right)  \left[  \widehat{\mathbf{M}}^{\left[  3\right]  }\left(
\mathsf{G}\right)  \right]  , \label{mrg}%
\end{equation}
which follows from the sum of formal products $\sum_{\mathsf{g}\in\mathsf{G}%
}r_{\mathsf{g}}\bullet\mathsf{g}$ (\ref{ra}).

\section{Public-key cryptosystems}

Public-key cryptosystems revolutionized cryptography by eliminating the need
for a shared secret key, instead utilizing a pair of mathematically linked
keys---a public key for encryption and a private key for decryption. We first
begin by introducing the foundational algorithm that made this paradigm
possible, the ElGamal cryptosystem~\cite{elg1985} which is based directly on
Diffie-Hellman ideas \cite{dif/hel1976}, but extends them into a full
asymmetric encryption scheme and a digital signature variant. Unlike RSA
\cite{riv/sha/adl1978}, which relies on the difficulty of integer
factorization, ElGamal's security is grounded in the computational
infeasibility of the Discrete Logarithm Problem in cyclic groups
\cite{odl1985,mau/wol1994}, most commonly implemented over prime fields or
elliptic curves~\cite{men/oor/van1996}. This probabilistic encryption
mechanism produces distinct ciphertexts for the same plaintext each time it is
run---a property that provides semantic security against chosen-plaintext
attacks when used with appropriate group parameters~\cite{kat/lin2020}. The
algorithm's influence persists today in standards such as the Digital
Signature Algorithm (DSA)~\cite{nist2013}, which is directly derived from
ElGamal signatures \cite{cha/cha/cha2003}, and in many privacy-preserving
cryptographic protocols~\cite{sti/pat2018}. There are many extensions and
improvements of the ElGamal cryptosystem, see, e.g.,
\cite{kar/lam2016,oka/dav/mam1998}, but all of them used only binary algebraic
structures. Here we generalize the ElGamal cryptosystem to ternary algebraic
structures (see, e.g., \cite{duplij2022}), while further its extensions can be
provided similarly to the binary case.

\subsection{Binary cryptosystem}

Let us formulate the abstract universal scheme of the binary cryptosystem
(which is for cyclic binary groups indeed the ElGamal one \cite{elg1985}) in a
unified way for binary groups $\mathsf{G}$, rings $\mathcal{R}$ and group
rings $\mathcal{R}[\mathsf{G}]$, and denote all them as $\mathit{M}$ with the
underlying set $M$. Their corresponding unit group is $\mathit{U}\left(
\mathit{M}\right)  $ with the underlying set $U_{M}$ of multiplicatively
invertible elements. For $u\in U_{M}$ and positive integer $n\geq0$, $u^{n}$
denotes $\left(  n-1\right)  $-fold composed binary multiplication (for cyclic
unit groups using $\operatorname{mod}$), we call it $n$-power, and negative
exponents are defined via the inverse $u^{-1}\in U_{M}$.

The general idea of the binary cryptosystem is to transfer a secret message
$\mathtt{m}\in M$ from \textit{Sender} to \textit{Recipient} by exchanging
public keys and cipher texts. The private (secret) key $\mathtt{x}$ is known
to \textit{Recipient} only, who cannot send it openly to \textit{Sender}, but
\textit{Recipient} can send openly some function $h=f\left(  \mathtt{x}%
\right)  $ which is difficult to solve (e.g., for cyclic groups this is the
Discrete Logarithm Problem \cite{odl1985,mau/wol1994}). Then \textit{Sender}
prepares as minimum two cipher texts (to make cancellation), such that the
private key $\mathtt{x}$ dependence can be cancelled out to give the searched
for secret message $\mathtt{m}$. The simplest choice for the function
$f\left(  \mathtt{x}\right)  $ is the exponential function, because the ratio
of two ciphers can be independent of $\mathtt{x}$.

The general algorithm of the binary cryptosystem consists of three steps:

\begin{enumerate}
\item \textbf{Key generation} by \textit{Recipient}.

\begin{itemize}
\item \textit{Recipient} takes an invertible (group unit) $u\in U_{M}$ which
is a public parameter, since $\mathit{M}=\mathsf{G},\mathcal{R},\mathcal{R}%
[\mathsf{G}]$ is known to all.

\item Using his own private integer key $\mathtt{x}$, \textit{Recipient}
prepares the public key using $\mathtt{x}$-powers%
\begin{equation}
h=u^{\mathtt{x}},\ \ \ \ \ \ \ \ h,u\in U_{M} \label{h}%
\end{equation}
and sends the pair $\left(  u,h\right)  $ to \textit{Sender}.
\end{itemize}

\item \textbf{Encryption} by \textit{Sender}.

\begin{itemize}
\item Having the public key $h$ and the secret message $\mathtt{m}$, for the
first glance \textit{Sender} can multiply them, but this will not hide
$\mathtt{m}$, the function should be more complicated.

\item \textit{Sender} chooses an additional parameter really to hide
$\mathtt{m}$, a random element (or ephemeral key), integer $\mathtt{k}$, that
is not known to \textit{Recipient} and exists locally, just for the purposes
of randomly encrypting a single secret message $\mathtt{m}$.

\item Then \textit{Sender} creates the following two-dimensional cipher text
by using $\mathtt{k}$-powers%
\begin{align}
c_{1}  &  =u^{\mathtt{k}},\label{c1}\\
c_{2}  &  =\mathtt{m}\cdot h^{\mathtt{k}}, \label{c2}%
\end{align}
where $\left(  \cdot\right)  $ is multiplication in the unit group
$\mathit{U}\left(  \mathit{M}\right)  $. Different random choices of
$\mathtt{k}$ give different two-dimensional ciphers $\left(  c_{1}%
,c_{2}\right)  $ which correspond to the same secret message $\mathtt{m}$.

\item \textit{Sender }sends the pair $\left(  c_{1},c_{2}\right)  $ to
\textit{Recipient} (but not the random element $\mathtt{k}$).
\end{itemize}

\item \textbf{Decryption} by \textit{Recipient}.

\begin{itemize}
\item \textit{Recipient} checks that the both parameters of two-dimension
cipher (\ref{c1})--(\ref{c2}) are in the unit group $\left(  c_{1}%
,c_{2}\right)  \in U_{M}\times U_{M}$ .

\item Then \textit{Recipient} uses his private key $\mathtt{x}$ and creates
the $\mathtt{x}$-power of the first cipher component $\left(  c_{1}\right)
^{\mathtt{x}}$, and check that it is in $U_{M}$ as well. The main idea of the
procedure is the $\mathtt{x}$-power ($\mathtt{x}$ is known to
\textit{Recipient} only) of the public cipher is derived through the public
parameter $h$, because, see (\ref{h}) and (\ref{c1}),%
\begin{equation}
\left(  c_{1}\right)  ^{\mathtt{x}}=\left(  u^{\mathtt{k}}\right)
^{\mathtt{x}}=\left(  u^{\mathtt{x}}\right)  ^{\mathtt{k}}=h^{\mathtt{k}}.
\label{cu}%
\end{equation}

\item The fraction $\mathtt{m}^{\prime}=c_{2}\left(  c_{1}\right)
^{-\mathtt{x}}$ (which exists, because $c_{1}\in U_{M}$) immediately gives the
sought for secret message $\mathtt{m}$. Indeed, using (\ref{h}) and
(\ref{c1})--(\ref{c2}), but without knowledge of the random element
$\mathtt{k}$, \textit{Recipient} obtains%
\begin{equation}
\mathtt{m}^{\prime}=c_{2}\left(  c_{1}\right)  ^{-\mathtt{x}}=\mathtt{m}\cdot
h^{\mathtt{k}}\cdot\left(  u^{\mathtt{k}}\right)  ^{-\mathtt{x}}%
=\mathtt{m}\cdot\left(  u^{\mathtt{x}}\right)  ^{\mathtt{k}}\cdot\left(
u^{\mathtt{k}}\right)  ^{-\mathtt{x}}\equiv\mathtt{m}. \label{cc}%
\end{equation}

Note, that \textit{Attacker}, the passive eavesdropper, can see all messages
sent over the public channel---the public parameters $u$, $\mathit{U}\left(
\mathit{M}\right)  $, the \textit{Recipient}'s public key $h$, and the
\textit{Sender}'s ciphertext $\left(  c_{1},c_{2}\right)  $---but cannot
decrypt and get $\mathtt{m}$, because \textit{Attacker} does not know the
\textit{Recipient}'s private key $\mathtt{x}$. Trying to solve the public key
definition (\ref{h}) (to get the private key $\mathtt{x}$) is the nontrivial
Discrete Logarithm Problem \cite{odl1985,mau/wol1994} and is related to the
Diffie-Hellman problem \cite{dif/hel1976}.
\end{itemize}
\end{enumerate}

\begin{example}
Consider the cyclic group $\mathsf{G}=\mathbb{Z}_{p}^{\times}=\mathbb{Z}%
_{23}^{\times}\equiv\mathbb{Z}\diagup23\mathbb{Z}=C_{11}$ of prime order 11
generated by $2$%
\begin{equation}
G=\langle2\rangle=\{1,2,4,8,16,9,18,13,3,6,12\}. \label{g}%
\end{equation}

\begin{enumerate}
\item \textbf{Key generation} by \textit{Recipient}. Takes $u=2$ and the
private integer key from the range $1\leq\mathtt{x}\leq10=p-1$, as
$\mathtt{x=7}$, and computes the public key (\ref{h}) by $h=2^{7}%
\operatorname{mod}23=13$, checking $13\in G$ (\ref{g}). \textit{Attacker} sees
the public parameters $\mathbb{Z}_{23}^{\times}$, $u=2$, and the public key
$h=13$, however, he cannot determine the private key $\mathtt{x=7}$ because
solving $2^{x}\equiv13\operatorname{mod}23$ is the Discrete Logarithm Problem
\cite{odl1985,mau/wol1994}. For the small parameters in this example, this is
easy to brute force. However, in a real cryptosystem, $p$ would be a prime of
size $2048$ bits or more, making the discrete logarithm problem
computationally infeasible.

\textit{Recipient} sends to \textit{Sender} the public pair $\left(
u,h\right)  =\left(  2,13\right)  $.

\item \textbf{Encryption} by \textit{Sender}. Wants to send \textit{Recipient}
the secret message $\mathtt{m=9}$, verifying that $\mathtt{m}\in G$, indeed
$\mathtt{9}=2^{5}\operatorname{mod}23\in G$ (\ref{h}). Then \textit{Sender}
chooses a random ephemeral key $\mathtt{k=4}$ ($1\leq\mathtt{k}\leq10$) and
creates two-dimensional cipher text (\ref{c1})--(\ref{c2}), by using
$\mathtt{k}$-powers of the received pair $\left(  u^{\mathtt{k}}%
,h^{\mathtt{k}}\right)  =\left(  2^{\mathtt{4}}\operatorname{mod}%
23,13^{\mathtt{4}}\operatorname{mod}23\right)  =\left(  16,18\right)  $,
proving $\left(  16,18\right)  \in G$, as $\left(  c_{1},c_{2}\right)
=\left(  u^{\mathtt{k}},\mathtt{m}\cdot h^{\mathtt{k}}\right)  =\left(
16,\left(  \mathtt{9}\cdot18\right)  \operatorname{mod}23\right)  =\left(
16,1\right)  $. \textit{Attacker} sees the ciphertext $(c_{1},c_{2})=(16,1)$
on the public channel, but without the private key $\mathtt{x=7}$, he cannot
recover the message $\mathtt{m=9}$ because that would require computing
$u^{-\mathtt{x}}=16^{-7}$, which needs knowledge of $\mathtt{x}$.

\textit{Sender} sends to \textit{Recipient} the public pair, two-dimensional
cipher $(c_{1},c_{2})=(16,1)$.

\item \textbf{Decryption} by \textit{Recipient}. Uses his private key
$\mathtt{x=7}$ and creates the $\mathtt{x}$-power of the first cipher
component $\left(  c_{1}\right)  ^{\mathtt{x}}=16^{\mathtt{7}}%
\operatorname{mod}23=2^{28}\operatorname{mod}23$, because the order of $G$ is
$11$, therefore in the power $28=28\operatorname{mod}11=6$, and
$2^{28\operatorname{mod}11}\operatorname{mod}23=2^{6}\operatorname{mod}%
23=18=\left(  c_{1}\right)  ^{\mathtt{x}}$, and \textit{Recipient }checks that
$18\in G$ as well. Computing the inverse by $18\cdot\left(  c_{1}\right)
^{-\mathtt{x}}=1\operatorname{mod}23$, he obtains $\left(  c_{1}\right)
^{-\mathtt{x}}=9$, and recovers the secret message by (\ref{cc}) as
$\mathtt{m}=\left(  1\cdot9\right)  \operatorname{mod}23=\mathtt{9}$.
\end{enumerate}
\end{example}

In the classical binary cryptosystem algorithms, the unit group $\mathit{U}%
\left(  \mathit{M}\right)  $ is assumed cyclic for several reasons: 1)
\textit{Uniform message space}. Every group element can be represented as
$u^{n}$ for some exponent $n$. 2) \textit{Hardness of the Discrete Logarithm
Problem} \cite{odl1985,mau/wol1994}, such that the best known algorithms
assume knowledge of the group order and its structure, while cyclic groups of
prime order are optimal. 3) During key generation, any element of group can be
a public key $h=u^{x}$.

However, these are cryptographic conveniences, not algebraic necessities.
Moreover, the main relation (\ref{cc}) of the above algorithm demands $\left(
u^{\mathtt{x}}\right)  ^{\mathtt{k}}\cdot\left(  u^{\mathtt{k}}\right)
^{-\mathtt{x}}=u^{\mathtt{xk}}\cdot u^{-\mathtt{kx}}=1$, $u\in U_{M}$. This
relation holds valid as for noncyclic groups, and even for nonabelian
(noncommutative) groups, since in both cases integer powers $r,s\in
\mathbb{N}_{0}$ of the same element commute $u^{r}{}^{s}=u^{sr}$. The only
algebraic requirement is that $u$ belongs to the unit group $U_{M}$ (so that
inverses $u^{-1}\in U_{M}$ exist) and that multiplication in the unit group
$\mathit{U}\left(  \mathit{M}\right)  $ is associative (for more complicated
nonassociative cryptosystem algorithms, see, e.g., \cite{che2022}).

\subsection{Ternary cryptosystem}

Now we propose a fundamentally novel approach to cryptosystems, which is based
on the new structures, ternary algebraic structures, see, e.g.,
\cite{duplij2022}. Here we present the abstract universal scheme of the
ternary cryptosystem generalizing the cyclic binary one
\cite{elg1985,hof/pip/sil2014} in a unified way for ternary groups
$\mathsf{G}^{\left[  3\right]  }$, rings with ternary product $\mathcal{R}%
^{\left[  m,3\right]  }$ and group rings $\mathcal{R}^{\left[  m,3\right]
}[\mathsf{G}^{\left[  3\right]  }]$, and denote all them as $\mathit{M}%
^{\left[  3\right]  }$ with the underlying set $M$. Their corresponding unit
group is $\mathit{U}\left(  \mathit{M}^{\left[  3\right]  }\right)  $ with the
underlying set $U_{M}$ of multiplicatively ternary invertible elements, such
that each element has its querelement (\ref{q}). For $u\in U_{M}$ and positive
integer $\ell\geq0$, $u^{2\ell+1}$ denotes $\ell$-fold composed ternary
multiplication (for cyclic unit groups using $\operatorname{mod}$), we call it
$\ell$-ternary power (\ref{xl}). For numbers we omit the symbol of product
$\mu$. The negative exponents of binary case correspond now to querelements
$u^{-1}\rightarrow Q\left(  u\right)  \in U_{M}$ (\ref{q}).

If we use the matrix ternarization, $\mathit{U}\left(  \mathit{B}\right)
\rightarrow\widehat{\mathit{U}}\left(  \mathit{M}^{\left[  3\right]  }\right)
$ and for the underlying set $U_{M}\rightarrow\widehat{U_{M}}$ (\ref{f}), to
obtain nonderived ternary groups, the underlying set of the unit ternary group
$U_{M}$ consists of the symbolic matrices of the form (\ref{mbb}), such that
each element of the matrix ternary unit group $\widehat{\mathit{U}}\left(
\mathit{M}^{\left[  3\right]  }\right)  $ is effectively a pair $\widehat
{u}=\left(  u_{\left(  1\right)  },u_{\left(  2\right)  }\right)  $, because%
\begin{equation}
\widehat{u}=\left(
\begin{array}
[c]{cc}%
0 & u_{\left(  1\right)  }\\
u_{\left(  2\right)  } & 0
\end{array}
\right)  . \label{u}%
\end{equation}

The goal of the ternary cryptosystem remains the same: to send from
\textit{Sender} to \textit{Recipient} the message $\mathtt{m}$ securely by
exchanging public keys and cipher texts only. The difference is in more
complicates structures of products, because they can be only of admissible
length (\ref{wl}). This make much more difficult to attack and find solutions
without knowledge of the private key which is now $\ell_{\mathtt{x}}$-ternary power.

The ternary cryptosystem algorithm consists of three steps:

\begin{enumerate}
\item \textbf{Key generation} by \textit{Recipient}.

\begin{itemize}
\item \textit{Recipient} takes an invertible, i.e. having querelement
(\ref{q}) (group unit) $\widehat{u}\in\widehat{U}_{M}$ which is a public
parameter (pair), since $\mathit{M}=\mathsf{G}^{\left[  3\right]
},\mathcal{R}^{\left[  m,3\right]  },\mathcal{R}^{\left[  m,3\right]
}[\mathsf{G}^{\left[  3\right]  }]$ is public, i.e. known to all.

\item \textit{Recipient} uses the private secret integer key $\ell
_{\mathtt{x}}$ and prepares the public key using $\ell_{\mathtt{x}}$-ternary
powers (below we omit the product symbol $\mu$)%
\begin{equation}
\widehat{h}=\mu^{\left[  3\right]  \circ\ell_{\mathtt{x}}}\left[  \widehat
{u}^{2\ell_{\mathtt{x}}+1}\right]  \equiv\left[  \widehat{u}^{2\ell
_{\mathtt{x}}+1}\right]  _{\times},\ \ \ \ \ \ \ \ \widehat{h},\widehat{u}%
\in\widehat{U}_{M}. \label{hh}%
\end{equation}
where $\left[  \ \right]  _{\times}$ is ternary multiplication (and its
ternary powers) in the matrix ternary unit group $\widehat{\mathit{U}}\left(
\mathit{M}\right)  $.

\item \textit{Recipient }sends openly the public pair $\left(  \widehat
{u},\widehat{h}\right)  $ (effectively 4 parameters) to \textit{Sender}.
\end{itemize}

\item \textbf{Encryption} by \textit{Sender}.

\begin{itemize}
\item \textit{Sender} chooses a secret message $\mathtt{m}$. To hide
$\mathtt{m}$ he introduce an additional parameter, a random element (or
ephemeral key), integer $\ell_{\mathtt{k}}$, that is not known to
\textit{Recipient} and exists locally, just for the purposes of randomly
encrypting a single secret message $\mathtt{m}$.

\item Then \textit{Sender} creates the following two-dimensional cipher text
by using $\ell_{\mathtt{k}}$-ternary powers%
\begin{align}
\widehat{c}_{1}  &  =\left[  \widehat{u}^{2\ell_{\mathtt{k}}+1}\right]
_{\times},\label{c11}\\
\widehat{c}_{2}  &  =\left[  \widehat{\mathtt{m}},\widehat{h}^{2\ell
_{\mathtt{k}}+1},\widehat{h}^{2\ell_{\mathtt{k}}+1}\right]  _{\times},
\label{c22}%
\end{align}
Different random choices of $\ell_{\mathtt{k}}$ give different two-dimensional
ciphers $\left(  \widehat{c}_{1},\widehat{c}_{2}\right)  $ which correspond to
the same secret message $\mathtt{m}$.

\item \textit{Sender }sends the pair $\left(  \widehat{c}_{1},\widehat{c}%
_{2}\right)  $ to \textit{Recipient} (without the random element
$\ell_{\mathtt{k}}$).
\end{itemize}

\item \textbf{Decryption} by \textit{Recipient}.

\begin{itemize}
\item \textit{Recipient} checks that the both parameters of two-dimension
cipher (\ref{c1})--(\ref{c2}) are in the ternary unit group $\left(
\widehat{c}_{1},\widehat{c}_{2}\right)  \in\widehat{U}_{M}\times\widehat
{U}_{M}$ .

\item Then \textit{Recipient} uses his private key $\ell_{\mathtt{x}}$ and
creates the $\ell_{\mathtt{x}}$-ternary power of the first cipher component
$\left(  \widehat{c}_{1}\right)  ^{2\ell_{\mathtt{x}}+1}$, and check that it
is in $\widehat{U}_{M}$ as well. Then, he obtains%
\begin{equation}
\left(  \widehat{c}_{1}\right)  ^{2\ell_{\mathtt{x}}+1}=\left[  \left(
\left[  \widehat{u}^{2\ell_{\mathtt{k}}+1}\right]  _{\times}\right)
^{2\ell_{\mathtt{x}}+1}\right]  _{\times}=\left[  \left(  \left[  \widehat
{u}^{2\ell_{\mathtt{x}}+1}\right]  _{\times}\right)  ^{2\ell_{\mathtt{k}}%
+1}\right]  _{\times}=\left[  \widehat{h}^{2\ell_{\mathtt{k}}+1}\right]
_{\times}, \label{ch}%
\end{equation}
which is the main idea of the ternary procedure: the $\ell_{\mathtt{x}}%
$-ternary power of the first cipher is expressed through the public
$\widehat{h}$, as in (\ref{cu}).

\item In the matrix ternary unit group $\widehat{\mathit{U}}\left(
\mathit{M}\right)  $ the role of inverse plays the querelement (\ref{q}). We
compute, using (\ref{ch})%
\begin{equation}
Q\left(  \left(  \widehat{c}_{1}\right)  ^{2\ell_{\mathtt{x}}+1}\right)
=\left[  \widehat{h}^{-\left(  2\ell_{\mathtt{k}}+1\right)  }\right]
_{\times}, \label{qc}%
\end{equation}
where on the r.h.s. there are usual matrix inverses.

\item Now we construct the ternary analog of the fraction $c_{2}%
c_{1}^{-\mathtt{x}}$ (\ref{cc}) as%
\begin{equation}
\widehat{\mathtt{m}}^{\prime}=\left[  \widehat{c}_{2},Q\left(  \left(
\widehat{c}_{1}\right)  ^{2\ell_{\mathtt{x}}+1}\right)  ,Q\left(  \left(
\widehat{c}_{1}\right)  ^{2\ell_{\mathtt{x}}+1}\right)  \right]  _{\times},
\label{c21}%
\end{equation}
which should give the secret message $\widehat{\mathtt{m}}$. Indeed, even
without knowledge of the random element $\ell_{\mathtt{k}}$, using (\ref{hh}),
(\ref{qc}), (\ref{c21}) and (\ref{c11})--(\ref{c22}), \textit{Recipient}
derives%
\begin{align}
\widehat{\mathtt{m}}^{\prime}  &  =\left[  \widehat{c}_{2},Q\left(  \left(
\widehat{c}_{1}\right)  ^{2\ell_{\mathtt{x}}+1}\right)  ,Q\left(  \left(
\widehat{c}_{1}\right)  ^{2\ell_{\mathtt{x}}+1}\right)  \right]  _{\times
}\nonumber\\
&  =\left[  \left[  \widehat{\mathtt{m}},\widehat{h}^{2\ell_{\mathtt{k}}%
+1},\widehat{h}^{2\ell_{\mathtt{k}}+1}\right]  _{\times},\left[  \widehat
{h}^{-\left(  2\ell_{\mathtt{k}}+1\right)  }\right]  _{\times},\left[
\widehat{h}^{-\left(  2\ell_{\mathtt{k}}+1\right)  }\right]  _{\times}\right]
_{\times}=\widehat{\mathtt{m}}. \label{m1}%
\end{align}

\end{itemize}
\end{enumerate}

The first difference from the binary case is the nontrivial ternary formulas,
which can be not known to \textit{Attacker}. Second, one matrix ternary
encryption/decryption procedure can transfer double information, because the
plaintext (secret messages) consists of pairs $\widehat{\mathtt{m}}=\left(
\mathtt{m}_{\left(  1\right)  },\mathtt{m}_{\left(  2\right)  }\right)  $, see
(\ref{u}).

If we do not use the matrix ternarization (\ref{f}), all the above formulas
hold valid for the ternary unit group $\mathit{U}\left(  \mathit{M}^{\left[
3\right]  }\right)  $, where $\mathit{M}^{\left[  3\right]  }=\mathsf{G}%
^{\left[  3\right]  },\mathcal{R}^{\left[  m,3\right]  },\mathcal{R}^{\left[
m,3\right]  }[\mathsf{G}^{\left[  3\right]  }]$.

\section{Examples of ternary cryptosystem}

Here we first consider a ternary cryptosystem over fractions without matrix
ternarization, then matrix ternarization of the finite binary group ring
$\mathbb{Z}_{5}[C_{3}]$, and finally over the finite $\left(  6,3\right)
$-ring with $5$ elements $\mathbb{Z}_{5}^{\left[  6,3\right]  }$.

\begin{example}
To demonstrate the validity of the proposed encryption and decryption
framework within a non-classical algebraic structure, we provide a concrete
numerical verification using a ternary ring of fractions which is actually a
ternary field $\mathbb{F}_{z}^{\left[  3\right]  }$. Let the set of fractions
be defined as:
\[
z(k,l)=\frac{(1+2k)}{(1+2l)},\ \ \ \ k,l\in\mathbb{Z},\ \ z(k,l)\in
\mathbb{F}_{z}^{\left[  3\right]  }.
\]

Before validating the cryptographic routine, we first formally verify that the
set $\{z(k,l)\mid k,l\in\mathbb{Z}\}$ is closed under the ternary
multiplication operation. By evaluating the product of three arbitrary
elements from this set, we obtain%
\begin{equation}
z(k_{1},l_{1})\cdot z(k_{2},l_{2})\cdot z(k_{3},l_{3})=\frac{(1+2k_{1}%
)(1+2k_{2})(1+2k_{3})}{(1+2l_{1})(1+2l_{2})(1+2l_{3})}=\frac{1+2K}{1+2L},
\end{equation}

where the transformed parameters $K$ and $L$ are defined as:
\[%
\begin{split}
K  &  =\frac{(1+2k_{1})(1+2k_{2})(1+2k_{3})-1}{2}\\
L  &  =\frac{(1+2l_{1})(1+2l_{2})(1+2l_{3})-1}{2}%
\end{split}
\]

Since the product of odd integers is always odd, the terms $(1+2k_{1}%
)(1+2k_{2})(1+2k_{3})-1$ and $(1+2l_{1})(1+2l_{2})(1+2l_{3})-1$ are guaranteed
to be even. Consequently, both $K$ and $L$ are strictly integers
($K,L\in\mathbb{Z}$), which directly confirms that the system satisfies the
closure property under ternary multiplication.

\textit{Recipient} chooses the system generator, e.g.,
\[
u=z(2,3)=\frac{1+4}{1+6}=\frac{5}{7},
\]
selects the private key parameter $\ell_{\mathtt{x}}=5$.

Let the initial plaintext secret message to be transmitted by \textit{Sender}
be%
\[
\mathtt{m}=z(1,2)=\frac{1+2}{1+4}=\frac{3}{5}.
\]
\textit{Sender} selects a random integer $\ell_{\mathtt{k}}=3$ as the
ephemeral session parameter. \textit{Recipient} establishes the public key
component $h$ by computing the ternary $\ell_{\mathtt{x}}$-power of the
generator $u$%
\begin{equation}
h=u^{2\ell_{\mathtt{x}}+1}=\left(  \frac{5}{7}\right)  ^{2\times5+1}=\left(
\frac{5}{7}\right)  ^{11}. \label{h1}%
\end{equation}

The resulting public key pair $(u,h)$ is published and sent to \textit{Sender}%
, while $\ell_{\mathtt{x}}=5$ is kept strictly confidential as the private key.

To encrypt the plaintext $\mathtt{m}$ using \textit{Recipient}'s public
parameters $(u,h)$, Sender uses the random session integer $\ell_{\mathtt{k}%
}=3$ to generate a two-part ciphertext pair $(c_{1},c_{2})$.The ephemeral
ciphertext component $c_{1}$ is computed by evaluating the power of the
generator $u$ with respect to the session parameter $\ell_{\mathtt{k}}$%
\[
c_{1}=u^{2\ell_{\mathtt{k}}+1}=\left(  \frac{5}{7}\right)  ^{2\times
3+1}=\left(  \frac{5}{7}\right)  ^{7}.
\]

Next, \textit{Sender} calculates the component by raising the public key $h$,
which is subsequently squared and multiplied by the plaintext message
$\mathtt{m}$ to produce the second ciphertext component%
\begin{equation}
c_{2}=m\cdot\left(  h^{2\ell_{\mathtt{k}}+1}\right)  ^{2}=\frac{3}{5}%
\cdot\left[  \left(  \frac{5}{7}\right)  ^{11}\right]  ^{7}=\frac{3}{5}%
\cdot\frac{5^{154}}{7^{154}}.
\end{equation}

Consequently, the complete ciphertext pair is given by:
\begin{equation}
(c_{1},c_{2})=\left(  \frac{5^{7}}{7^{7}},\frac{3}{5}\cdot\frac{5^{154}%
}{7^{154}}\right)
\end{equation}

\textit{Recipient} then computes the algebraic querelement (\ref{q}) of the
first reconstructed component (see (\ref{h1}))%
\[
(c_{1})^{2\ell_{\mathtt{x}}+1}=\left(  \frac{5}{7}\right)  ^{7\times11}%
=\frac{5^{77}}{7^{77}}=h^{2\ell_{\mathtt{k}}+1},
\]%
\[
Q\left(  (c_{1})^{2\ell_{\mathtt{x}}+1}\right)  =\left(  (c_{1})^{2\ell
_{\mathtt{k}}+1}\right)  ^{-1}=\frac{7^{77}}{5^{77}}.
\]

Finally, the original plaintext message $m^{\prime}$ is fully recovered by
multiplying the ciphertext component $c_{2}$ with the square of the computed
inverse%
\begin{equation}
\mathtt{m}^{\prime}=c_{2}\cdot\left[  Q\left(  (c_{1})^{2\ell_{\mathtt{x}}%
+1}\right)  \right]  ^{2}=\frac{3}{5}\cdot\frac{5^{154}}{7^{154}}\cdot\left(
\frac{7^{77}}{5^{77}}\right)  ^{2}=\frac{3}{5}\equiv\mathtt{m}.
\end{equation}

Such calculation of fractions is possible, because their transferring can be
provided without direct divisions, using only ordinary integers.
\end{example}

\begin{example}
\label{exam-z5c3}Consider the ternary cryptographic procedure implemented
within a matrix ternary group ring structure made by the matrix ternarization
(\ref{mrg}) of $\mathbb{Z}_{5}[\mathsf{C}_{3}]$, where the cyclic group has
order $3$, such that $\mathsf{C}_{3}=\left\{  e,\omega,\omega^{2}\right\}  $,
$\omega^{3}=e$. The product in $\mathsf{C}_{3}$ is%
\begin{equation}
\omega^{i}\omega^{j}=\omega^{\left(  i+j\right)  \operatorname{mod}3}%
,i,j\in\mathbb{N}_{0}, \label{w}%
\end{equation}
while for general elements $\mathsf{g},\mathsf{h}\in\mathsf{C}_{3}$ we use for
their modular product the corresponding to (\ref{w}) notation $\left(
\mathsf{gh}\right)  \widehat{\operatorname{mod}}3$, informally. The ternary
products $\left[  \ \ \right]  $ are now triple symbolic matrix products.
Informally, using the symbolic matrix representation (\ref{mbb}), we write%
\begin{align}
\widehat{\mathrm{R}}^{\left[  \mathbf{2},\mathbf{3}\right]  }  &
=\widehat{\mathbf{M}}^{\left[  \mathbf{3}\right]  }\left(  \mathbb{Z}%
_{5}[\mathsf{C}_{3}]\right)  =\widehat{\mathbf{M}}^{\left[  3\right]  }\left(
\mathbb{Z}_{5}\right)  \left[  \widehat{\mathbf{M}}^{\left[  3\right]
}\left(  \mathsf{C}_{3}\right)  \right] \nonumber\\
&  \Rightarrow\left(
\begin{array}
[c]{cc}%
0 & \mathbb{Z}_{5}[\mathsf{C}_{3}]\\
\mathbb{Z}_{5}[\mathsf{C}_{3}] & 0
\end{array}
\right)  =\left(
\begin{array}
[c]{cc}%
0 & \mathbb{Z}_{5}\\
\mathbb{Z}_{5} & 0
\end{array}
\right)  \otimes\left(
\begin{array}
[c]{cc}%
0 & \mathsf{C}_{3}\\
\mathsf{C}_{3} & 0
\end{array}
\right)  ,
\end{align}
such that an element of the ternary group ring $\widehat{\mathrm{R}}^{\left[
\mathbf{2},\mathbf{3}\right]  }=\widehat{\mathbf{M}}^{\left[  \mathbf{3}%
\right]  }\left(  \mathbb{Z}_{5}[\mathsf{C}_{3}]\right)  $ can be presented as
a formal sum of the direct product of pairs (\ref{ra})%
\begin{align}
\widehat{\mathrm{r}}  &  =%
%TCIMACRO{\dsum \limits_{\mathsf{g}}}%
%BeginExpansion
{\displaystyle\sum\limits_{\mathsf{g}}}
%EndExpansion
\left(
\begin{array}
[c]{cc}%
0 & r_{g\left(  1\right)  }\\
r_{g\left(  2\right)  } & 0
\end{array}
\right)  \otimes\left(
\begin{array}
[c]{cc}%
0 & \mathsf{g}_{\left(  1\right)  }\\
\mathsf{g}_{\left(  2\right)  } & 0
\end{array}
\right) \nonumber\\
&  \Rightarrow%
%TCIMACRO{\dsum \limits_{\mathsf{g}}}%
%BeginExpansion
{\displaystyle\sum\limits_{\mathsf{g}}}
%EndExpansion
\left(  r_{g\left(  1\right)  },r_{g\left(  2\right)  }\right)  \otimes\left(
\mathsf{g}_{\left(  1\right)  },\mathsf{g}_{\left(  2\right)  }\right)
,\ \ r_{g\left(  1,2\right)  }\in\mathbb{Z}_{5},\ \ \mathsf{g}_{\left(
1,2\right)  }\in\mathsf{C}_{3},\ \ \widehat{\mathrm{r}}\in\widehat{\mathrm{R}%
}^{\left[  \mathbf{2},\mathbf{3}\right]  }. \label{rr}%
\end{align}
The ternary product in the group ring $\widehat{\mathrm{R}}^{\left[
\mathbf{2},\mathbf{3}\right]  }$ in the manifest form is (following from
(\ref{rr}))%
\begin{align}
\left[  \widehat{\mathrm{r}}^{\prime},\widehat{\mathrm{r}}^{\prime\prime
},\widehat{\mathrm{r}}^{\prime\prime\prime}\right]   &  =%
%TCIMACRO{\dsum \limits_{\mathsf{h}}}%
%BeginExpansion
{\displaystyle\sum\limits_{\mathsf{h}}}
%EndExpansion
\left(
\begin{array}
[c]{cc}%
0 & r_{g\left(  1\right)  }^{\prime}r_{g\left(  2\right)  }^{\prime\prime
}r_{g\left(  1\right)  }^{\prime\prime\prime}\\
r_{g\left(  2\right)  }^{\prime}r_{g\left(  1\right)  }^{\prime\prime
}r_{g\left(  2\right)  }^{\prime\prime\prime} & 0
\end{array}
\right)  \otimes\left(
\begin{array}
[c]{cc}%
0 & \mathsf{h}_{\left(  1\right)  }=\mathsf{g}_{\left(  1\right)  }^{\prime
}\mathsf{g}_{\left(  2\right)  }^{\prime\prime}\mathsf{g}_{\left(  1\right)
}^{\prime\prime\prime}\\
\mathsf{h}_{\left(  2\right)  }=\mathsf{g}_{\left(  2\right)  }^{\prime
}\mathsf{g}_{\left(  1\right)  }^{\prime\prime}\mathsf{g}_{\left(  2\right)
}^{\prime\prime\prime} & 0
\end{array}
\right) \nonumber\\
&  \Rightarrow%
%TCIMACRO{\dsum \limits_{\mathsf{g}}}%
%BeginExpansion
{\displaystyle\sum\limits_{\mathsf{g}}}
%EndExpansion
\left(  r_{g\left(  1\right)  }^{\prime}r_{g\left(  2\right)  }^{\prime\prime
}r_{g\left(  1\right)  }^{\prime\prime\prime},r_{g\left(  2\right)  }^{\prime
}r_{g\left(  1\right)  }^{\prime\prime}r_{g\left(  2\right)  }^{\prime
\prime\prime}\right)  \operatorname{mod}5\otimes\left(  \mathsf{h}_{\left(
1\right)  },\mathsf{h}_{\left(  2\right)  }\right)  \widehat
{\operatorname{mod}}3.
\end{align}
All scalar operations are evaluated $\operatorname{mod}5$ within the finite
field $\mathbb{Z}_{5}$, and the group components are calculated $\widehat
{\operatorname{mod}}3$, respecting the cyclic structure of $\mathsf{C}_{3}$
(\ref{w}). The unity of the group ring $\widehat{\mathrm{R}}^{\left[
\mathbf{2},\mathbf{3}\right]  }$ is%
\begin{equation}
\widehat{\mathrm{e}}=\left(  1,1\right)  \otimes\left(  e,e\right)
,\ \ \ \ \ \ 1\in\mathbb{Z}_{5},\ \ \ e\in\mathsf{C}_{3}. \label{e1}%
\end{equation}

Let the initial plaintext message to be transmitted be $\widehat{\mathtt{m}%
}=(4,3)\otimes(\omega,\omega^{2})$. \textit{Recipient} selects an integer
$\ell_{\mathtt{x}}=2$ to serve as the private key, while \textit{Sender}
chooses a random integer $\ell_{\mathtt{k}}=3$ as the ephemeral parameter for
the encryption process. Furthermore, we select a ternary invertible element
$\widehat{u}=(2,2)\otimes(\omega,e)$ that possesses a unique querelement
(\ref{q}). For simplicity, we take for $\widehat{\mathtt{m}}$ and $\widehat
{u}$ the monomial elements of the group ring $\widehat{\mathrm{R}}^{\left[
\mathbf{2},\mathbf{3}\right]  }$ (\ref{rr}).

The querelement $Q\left(  \widehat{u}\right)  $ associated with the base
element $\widehat{u}$ must satisfy the fundamental ternary identity $\left[
\widehat{u},\widehat{u},Q\left(  \widehat{u}\right)  \right]  =u$. To
determine the explicit components of $Q\left(  \widehat{u}\right)  $, we
assume it takes the general form
\begin{equation}
Q\left(  \widehat{u}\right)  =\left(  a,b\right)  \otimes\left(  w^{c}%
,w^{d}\right)  \label{qa}%
\end{equation}

Substituting $\widehat{u}=(2,2)\otimes(\omega,e)$ and the assumed structure of
$Q\left(  \widehat{u}\right)  $ into the ternary product identity yields the
following evaluation%
\begin{equation}
\left[  \widehat{u},\widehat{u},Q\left(  \widehat{u}\right)  \right]
=(4a,4b)\otimes(\omega^{c+1},\omega^{d})=(2,2)\otimes(\omega,e)
\end{equation}
By equating the corresponding components on both sides of the equation, we
derive the system of modular relations over $\mathbb{Z}_{5}$ and $C_{3}$, and
solving this system yields the querelement of $\widehat{u}$ within this
ternary group ring structure is explicitly determined as%
\begin{equation}
Q\left(  \widehat{u}\right)  =(3,3)\otimes(e,\omega). \label{q3}%
\end{equation}

To establish the cyclic properties of the generator under ternary
multiplication, we sequentially evaluate the higher-order ternary powers of
$\widehat{u}=(2,2)\otimes(\omega,e)$ using modulo $5$ for the ring part, and
modulo $3$ for the group part and the recurrent formula $\widehat{u}%
^{n+2}=[\widehat{u}^{n},\widehat{u},\widehat{u}]$.%
\begin{equation}%
\begin{split}
\widehat{u}^{3}  &  =(8,8)\operatorname{mod}5\otimes(\omega^{3},\omega
)\widehat{\operatorname{mod}}3=(3,3)\otimes(e,\omega)\\
\widehat{u}^{5}  &  =(12,12)\operatorname{mod}5\otimes(e,\omega^{2}%
)\widehat{\operatorname{mod}}3=(2,2)\otimes(e,\omega^{2})\\
\widehat{u}^{7}  &  =(8,8)\operatorname{mod}5\otimes(\omega,\omega
^{3})\widehat{\operatorname{mod}}3=(3,3)\otimes(\omega,e)\\
\widehat{u}^{9}  &  =(12,12)\operatorname{mod}5\otimes(\omega^{2}%
,\omega)\widehat{\operatorname{mod}}3=(2,2)\otimes(\omega^{2},\omega)\\
\widehat{u}^{11}  &  =(8,8)\operatorname{mod}5\otimes(e,\omega^{2}%
)\widehat{\operatorname{mod}}3=(3,3)\otimes(e,\omega^{2})\\
\widehat{u}^{13}  &  =(12,12)\operatorname{mod}5\otimes(\omega,\omega
^{3})\widehat{\operatorname{mod}}3=(2,2)\otimes(\omega,e)\equiv\widehat{u}.
\end{split}
\label{uu}%
\end{equation}

Since $\widehat{u}^{13}=\widehat{u}$, the algebraic sequence repeats,
confirming that the generator possesses a strict cycle length of $12$ within
this structure, such that%
\begin{equation}
\widehat{u}^{i}\widehat{u}^{j}\widehat{u}^{k}=\widehat{u}^{\left(
i+j+k\right)  \operatorname{mod}12},\ \ \ \ i,j,k\in\mathbb{N}_{0}%
,\ \ i,j,k\ \text{are odd}. \label{uuu}%
\end{equation}
Thus, the ternary subgroup of the group ring $\widehat{\mathrm{R}}^{\left[
\mathbf{2},\mathbf{3}\right]  }$ (\ref{rr}) generated by $\widehat
{u}=(2,2)\otimes(\omega,e)$ and unity (\ref{e1}) is the cyclic ternary group
$\mathsf{C}_{12}^{\left[  3\right]  }$ with multiplication satisfying
(\ref{uuu}), $\widehat{u}^{0}=\widehat{\mathrm{e}}$, and the unique
querelement $Q\left(  \widehat{u}^{i}\right)  =\widehat{u}^{\left(
12-i\right)  \operatorname{mod}12}$.

Based on the selected private key $\ell_{\mathtt{x}}=2$ and $\ell_{\mathtt{k}%
}=3$, the corresponding exponent dimensions for the cryptographic routine are
$2\ell_{\mathtt{x}}+1=5$ and $2\ell_{\mathtt{k}}+1=7$.

\textit{Recipient} establishes the public key component $\widehat{h}$ by
\[
\widehat{h}=\widehat{u}^{2\ell_{\mathtt{x}}+1}=\widehat{u}^{5}=(2,2)\otimes
(e,\omega^{2})
\]

To encrypt the plaintext message $\mathtt{m}$, \textit{Sender} computes a
ciphertext pair $(\widehat{c}_{1},\widehat{c}_{2})$. The ciphertext components
$\widehat{c}_{1}$ and $\widehat{c}_{2}$ are generated as%
\[
\widehat{c}_{1}=\widehat{u}^{2\ell_{\mathtt{k}}+1}=\widehat{u}^{7}%
=(3,3)\otimes(\omega,e).
\]%
\begin{equation}
c_{2}=[\widehat{\mathtt{m}},\widehat{h}^{2l_{k}+1},\widehat{h}^{2l_{k}%
+1}]=[\widehat{\mathtt{m}},\widehat{u}^{35},\widehat{u}^{35}].
\end{equation}

The exponent can be reduced via modular reduction yielding
$u^{35\operatorname{mod}12}=u^{11}$ (see (\ref{uuu})).
\[%
\begin{split}
\widehat{c}_{2}  &  =[(4,3)\otimes(\omega,\omega^{2}),(3,3)\otimes
(e,\omega^{2}),(3,3)\otimes(e,\omega^{2})]\\
&  =(36,27)\operatorname{mod}5\otimes(\omega^{3},\omega^{4})\widehat
{\operatorname{mod}}3=(1,2)\otimes(e,\omega).
\end{split}
\]

Thus, the completed ciphertext pair is given by%
\[
(\widehat{c}_{1},\widehat{c}_{2})=\left(  (3,3)\otimes(\omega,e),(1,2)\otimes
(e,\omega)\right)  .
\]

Upon receiving the ciphertext pair $(\widehat{c}_{1},\widehat{c}_{2})$,
\textit{Recipient} initiates the decryption process (see (\ref{uu}) and
(\ref{uuu}))
\[
(c_{1})^{2\ell_{\mathtt{x}}+1}=(u^{7})^{5}=u^{35\operatorname{mod}12}%
=u^{11}=(3,3)\otimes(e,\omega^{2})
\]

The recipient then computes the querelement of this recovered component
\[
Q((\widehat{c}_{1})^{2\ell_{\mathtt{x}}+1})=(a,b)\otimes(\omega^{c},\omega
^{d})
\]

By the defining ternary relation and using (\ref{q}), we establish that
\[
\lbrack(\widehat{c}_{1})^{2\ell_{\mathtt{x}}+1},(\widehat{c}_{1}%
)^{2\ell_{\mathtt{x}}+1},Q((\widehat{c}_{1})^{2\ell_{\mathtt{x}}%
+1})]=(\widehat{c}_{1})^{2\ell_{\mathtt{x}}+1}%
\]

Substituting the known values yields the following algebraic expansion:
\[%
\begin{split}
\lbrack(3,3)\otimes(e,\omega^{2}),(3,3)\otimes(e,\omega^{2}),(a,b)\otimes
(\omega^{c},\omega^{d})]  &  =(9a,9b)\otimes(\omega^{2+c},\omega^{2+d})\\
&  =(3,3)\otimes(e,\omega^{2}).
\end{split}
\]

Solving the above modular equation with respects to $\left(  a,b,c,d\right)  $
yields the precise operator value of the querelement
\[
Q((\widehat{c}_{1})^{2\ell_{\mathtt{x}}+1})=(2,2)\otimes(\omega,e).
\]

The original plaintext is fully reconstructed by executing a ternary
multiplication of $c_{2}$ with two instances of the computed querelement%
\begin{align}
\mathtt{m}^{\prime}  &  =[\widehat{c}_{2},Q((\widehat{c}_{1})^{2\ell
_{\mathtt{x}}+1}),Q((\widehat{c}_{1})^{2\ell_{\mathtt{x}}+1})]=[(1,2)\otimes
(e,\omega),(2,2)\otimes(\omega,e),(2,2)\otimes(\omega,e)]\nonumber\\
&  =(4,8)\operatorname{mod}5\otimes(\omega,\omega^{2})\widehat
{\operatorname{mod}}3=(4,3)\otimes(\omega,\omega^{2})=\mathtt{m}%
\end{align}

The resulting decryption yields $m^{\prime}=m$, which perfectly matches the
initial plaintext message. The dimension of any monomial element of the matrix
ternary group ring $\widehat{\mathrm{R}}^{\left[  \mathbf{2},\mathbf{3}%
\right]  }$ is four, see (\ref{rr}). This gives minimal number of plaintext
parameters, which can be transferred. If the secret message $\mathtt{m}$
consists of $N$ formal summands in (\ref{rr}), the ternary cryptosystem
algorithm can transfer $4N$ plaintext parameters at once.
\end{example}

Now we consider the finite matrix ternary group ring, when the ring part is
matrix ternarization of the finite $\left(  m,3\right)  $-ring, but not a
binary ring, as in \textit{Example} \ref{exam-z5c3}. Recall, that the finite
$\left(  m,n\right)  $-ring is the set of secondary congruence classes of
infinite $\left(  m,n\right)  $-ring $\mathbb{Z}^{\left[  a,b\right]  }(m,n)$
being set of representatives of the congruence class $\left[  a\right]  _{b}$,
$0\leq a\leq b-1$, $b\in\mathbb{N}$. That is the finite $\left(  m,n\right)
$-ring $\mathbb{Z}_{q}^{\left[  a,b\right]  }(m,n)$ having $q$ elements (of
order $q$) is defined by $\mathbb{Z}_{q}^{\left[  a,b\right]  }%
(m,n)=\mathbb{Z}^{\left[  a,b\right]  }(m,n)\diagup\left(  bq\right)
\mathbb{Z}$. The corresponding exotic $\left(  m,n\right)  $-fields have
unusual properties, and can be zeroless, zeroless-nonunital or have several
units, and it is even possible for all elements to be units, also there exist
non-isomorphic finite fields of the same arity shape and order. None of the
above is possible in the binary case (for further details, see
\cite{dup2017a,duplij2022}).

\begin{example}
Consider the ternary structures over the finite $\left(  6,3\right)  $-ring
with $6$-ary addition and ternary multiplication $\mathcal{R}^{\left[
6,3\right]  }=\mathbb{Z}_{5}^{\left[  6,3\right]  }(4,5)$ of order $5$, with
the underlying set $R=\{4,9,14,19,24\}$, the defined ternary multiplication
operator $\left[  \ \ \right]  $, brackets as before. The calculations for the
ring components are performed modulo $bq=25$. The Cayley table is given in
TABLE 1. \begin{table}[ptbh]
\caption{Ternary multiplication table of the finite field $\mathbb{F}%
_{5}^{\left[  6,3\right]  }$ ($\operatorname{mod}25$)}%
\centering
\renewcommand{\arraystretch}{1.2}
\begin{tabular}
[c]{|c|c|c|c|c|c|}\hline
No. & Multiplication & No. & Multiplication & No. & Multiplication\\\hline
1 & $\left[  4,4,4\right]  =14$ & 2 & $\left[  9,9,9\right]  =4$ & 3 &
$\left[  14,14,14\right]  =19$\\\hline
4 & $\left[  19,19,19\right]  =9$ & 5 & $\left[  24,24,24\right]  =24$ & 6 &
$\left[  4,4,9\right]  =19$\\\hline
7 & $\left[  9,9,4\right]  =24$ & 8 & $\left[  4,4,14\right]  =24$ & 9 &
$\left[  14,14,4\right]  =9$\\\hline
10 & $\left[  4,4,19\right]  =4$ & 11 & $\left[  19,19,4\right]  =19$ & 12 &
$\left[  4,4,24\right]  =9$\\\hline
13 & $\left[  24,24,4\right]  =4$ & 14 & $\left[  9,9,14\right]  =9$ & 15 &
$\left[  14,14,9\right]  =14$\\\hline
16 & $\left[  9,9,19\right]  =14$ & 17 & $\left[  19,19,9\right]  =24$ & 18 &
$\left[  9,9,24\right]  =19$\\\hline
19 & $\left[  24,24,9\right]  =9$ & 20 & $\left[  14,14,19\right]  =24$ & 21 &
$\left[  19,19,14\right]  =4$\\\hline
22 & $\left[  14,14,24\right]  =4$ & 23 & $\left[  24,24,14\right]  =14$ &
24 & $\left[  19,19,24\right]  =14$\\\hline
25 & $\left[  24,24,19\right]  =19$ & 26 & $\left[  4,9,14\right]  =4$ & 27 &
$\left[  4,9,19\right]  =9$\\\hline
28 & $\left[  4,9,24\right]  =14$ & 29 & $\left[  4,14,19\right]  =14$ & 30 &
$\left[  4,14,24\right]  =19$\\\hline
31 & $\left[  4,19,24\right]  =24$ & 32 & $\left[  9,14,19\right]  =19$ & 33 &
$\left[  9,14,24\right]  =24$\\\hline
34 & $\left[  9,19,24\right]  =4$ & 35 & $\left[  14,19,24\right]  =9$ &  &
\\\hline
\end{tabular}
\end{table}

The total number of distinct ternary multiplications required to completely
define the algebraic operation can be computed via the following expression:
\[
C_{5}^{1}+C_{5}^{1}C_{4}^{1}+C_{5}^{3}=35.
\]

In the above equation, the first term $C_{5}^{1}$ accounts for the scenarios
where all three elements in the ternary product are identical; the second term
$C_{5}^{1}C_{4}^{1}$ represents the instances consisting of two identical
elements and one distinct element; the third term $C_{5}^{3}$ corresponds to
the configurations where all three elements are mutually distinct.

It follows from TABLE 1 that $24_{e}=24$ is only one ternary idempotent (from
No.$5$) and it is unity (from Nos. $13,19,23,25$). All other elements have a
unique querelement%
\begin{equation}
Q\left(  4\right)  \overset{No.10}{=}19,\ \ Q\left(  19\right)  \overset
{No.11}{=}4,\ \ \ \ \ Q\left(  9\right)  \overset{No.14}{=}14\ \ Q\left(
14\right)  \overset{No.15}{=}9,\ \ \ \ Q\left(  24\right)  \overset{No.5}%
{=}24,
\end{equation}
and therefore $\mathbb{Z}_{5}^{\left[  6,3\right]  }(4,5)$ is actually a
finite $\left(  6,3\right)  $-field $\mathbb{F}_{5}^{\left[  6,3\right]  }$ of
order $q=5$, such that the group of units (ternary invertible elements)
coincide with the whole underlying set.

The matrix ternarization is provided by analogy with \textit{Example}
\ref{exam-z5c3} with the same group part $\mathsf{C}_{3}$, where the cyclic
group is $\mathsf{C}_{3}=\left\{  e,\omega,\omega^{2}\right\}  $, $\omega
^{3}=e$. The symbolic matrix representation (\ref{mbb}) now becomes%
\begin{align}
\widehat{\mathrm{F}}^{\left[  \mathbf{6},\mathbf{3}\right]  }  &
=\widehat{\mathbf{M}}^{\left[  \mathbf{3}\right]  }\left(  \mathbb{F}%
_{5}^{\left[  6,3\right]  }[\mathsf{C}_{3}]\right)  =\widehat{\mathbf{M}%
}^{\left[  3\right]  }\left(  \mathbb{F}_{5}^{\left[  6,3\right]  }\right)
\left[  \widehat{\mathbf{M}}^{\left[  3\right]  }\left(  \mathsf{C}%
_{3}\right)  \right] \nonumber\\
&  \Rightarrow\left(
\begin{array}
[c]{cc}%
0 & \mathbb{F}_{5}^{\left[  6,3\right]  }[\mathsf{C}_{3}]\\
\mathbb{F}_{5}^{\left[  6,3\right]  }[\mathsf{C}_{3}] & 0
\end{array}
\right)  =\left(
\begin{array}
[c]{cc}%
0 & \mathbb{F}_{5}^{\left[  6,3\right]  }\\
\mathbb{F}_{5}^{\left[  6,3\right]  } & 0
\end{array}
\right)  \otimes\left(
\begin{array}
[c]{cc}%
0 & \mathsf{C}_{3}\\
\mathsf{C}_{3} & 0
\end{array}
\right)  , \label{f6}%
\end{align}
and all other notations are the same. Thus, the matrix group ring becomes also
$\left(  \mathbf{6,3}\right)  $-field $\widehat{\mathrm{F}}^{\left[
\mathbf{6},\mathbf{3}\right]  }$, because all elements (excluding zero) are
ternary invertible, and the $6$-ary addition is inherited from $\mathbb{Z}%
_{5}^{\left[  6,3\right]  }(4,5)$ by linearity. In this way, we can choose any
elements as generators for the corresponding ternary cryptosystem. The unity
of the group field $\widehat{\mathrm{F}}^{\left[  \mathbf{6},\mathbf{3}%
\right]  }$ is%
\begin{equation}
\widehat{\mathrm{e}}=\left(  24,24\right)  \otimes\left(  e,e\right)
,\ \ \ \ \ \ 24\in\mathbb{F}_{5}^{\left[  6,3\right]  },\ \ \ e\in
\mathsf{C}_{3}. \label{ee}%
\end{equation}

Let the plaintext message to be securely transmitted be $\mathtt{m}%
=(9,14)\otimes(\omega,\omega^{2})$. We select the system generator to be
$\widehat{u}=(4,9)\otimes(e,\omega)\in\widehat{\mathrm{F}}^{\left[
\mathbf{6},\mathbf{3}\right]  }$, the private key as $\ell_{\mathtt{x}}=2$,
and choose a random integer $\ell_{\mathtt{k}}=3$ as the ephemeral parameter
for the encryption.

We establish the existence and uniqueness of the querelement $Q\left(
\widehat{u}\right)  $ corresponding to the generator $u$, which must strictly
satisfy the fundamental ternary identity $\left[  \widehat{u},\widehat
{u},Q\left(  \widehat{u}\right)  \right]  =\widehat{u}$. Let the general form
of the target querelement be expressed as (see (\ref{qa})) $Q\left(
\widehat{u}\right)  =\left(  a,b\right)  \otimes\left(  w^{c},w^{d}\right)  $,
where $a,b,c,d\in\mathbb{Z}$. Expanding the ternary product yields%

\begin{align}
\left[  \widehat{u},\widehat{u},Q\left(  \widehat{u}\right)  \right]   &
=\left[  (4,9)\otimes(e,\omega),(4,9)\otimes(e,\omega),\left(  a,b\right)
\otimes\left(  w^{c},w^{d}\right)  \right] \nonumber\\
&  =(36a,36b)\otimes(\omega^{c+1},\omega^{d+1})=(4,9)\otimes(e,\omega).
\end{align}

Solving the modular equations (the ring part $\operatorname{mod}25$ and the
group part $\widehat{\operatorname{mod}}3$) gives the unique querelement%
\begin{equation}
Q\left(  \widehat{u}\right)  =(14,19)\otimes(\omega^{2},e).
\end{equation}

Periodicity analysis of the ternary powers of $\widehat{u}$ can be simplified
using the recurrent formula $\widehat{u}^{n+2}=[\widehat{u}^{n},\widehat
{u},\widehat{u}]$ as follows%

\begin{align}
\widehat{u}  &  =(4,9)\otimes(e,\omega)\nonumber\\
\widehat{u}^{3}  &  =[(4,9)\otimes(e,\omega),(4,9)\otimes(e,\omega
),(4,9)\otimes(e,\omega)]\nonumber\\
&  =(144,324)\pmod{25}\otimes(\omega,\omega^{2})\pmod{3}=(19,24)\otimes
(\omega,\omega^{2})\nonumber\\
\widehat{u}^{5}  &  =[\widehat{u}^{3},\widehat{u},\widehat{u}]=[(19,24)\otimes
(\omega,\omega^{2}),(4,9)\otimes(e,\omega),(4,9)\otimes(e,\omega)]\nonumber\\
&  =(684,864)\pmod{25}\otimes(\omega^{2},\omega^{3})\pmod{3}=(9,14)\otimes
(\omega^{2},e)\nonumber\\
\widehat{u}^{7}  &  =[\widehat{u}^{5},\widehat{u},\widehat{u}]=[(9,14)\otimes
(\omega^{2},e),(4,9)\otimes(e,\omega),(4,9)\otimes(e,\omega)]\nonumber\\
&  =(324,504)\pmod{25}\otimes(\omega^{3},\omega)\pmod{3}=(24,4)\otimes
(e,\omega)\nonumber
\end{align}%
\begin{align}
\widehat{u}^{9}  &  =[\widehat{u}^{7},\widehat{u},\widehat{u}]=[(24,4)\otimes
(e,\omega),(4,9)\otimes(e,\omega),(4,9)\otimes(e,\omega)]\nonumber\\
&  =(864,144)\pmod{25}\otimes(\omega,\omega^{2})\pmod{3}=(14,19)\otimes
(\omega,\omega^{2})\nonumber\\
\widehat{u}^{11}  &  =[\widehat{u}^{9},\widehat{u},\widehat{u}%
]=[(14,19)\otimes(\omega,\omega^{2}),(4,9)\otimes(e,\omega),(4,9)\otimes
(e,\omega)]\nonumber\\
&  =(504,684)\pmod{25}\otimes(\omega^{2},\omega^{3})\pmod{3}=(4,9)\otimes
(\omega^{2},e)\nonumber
\end{align}%
\begin{align}
\widehat{u}^{13}  &  =[\widehat{u}^{11},\widehat{u},\widehat{u}]=[(4,9)\otimes
(\omega^{2},e),(4,9)\otimes(e,\omega),(4,9)\otimes(e,\omega)]\nonumber\\
&  =(144,324)\pmod{25}\otimes(\omega^{3},\omega)\pmod{3}=(19,24)\otimes
(e,\omega)\nonumber\\
\widehat{u}^{15}  &  =[\widehat{u}^{13},\widehat{u},\widehat{u}%
]=[(19,24)\otimes(e,\omega),(4,9)\otimes(e,\omega),(4,9)\otimes(e,\omega
)]\nonumber\\
&  =(684,864)\pmod{25}\otimes(\omega,\omega^{2})\pmod{3}=(9,14)\otimes
(\omega,\omega^{2})\nonumber
\end{align}%
\begin{align}
\widehat{u}^{17}  &  =[\widehat{u}^{15},\widehat{u},\widehat{u}%
]=[(9,14)\otimes(\omega,\omega^{2}),(4,9)\otimes(e,\omega),(4,9)\otimes
(e,\omega)]\nonumber\\
&  =(324,504)\pmod{25}\otimes(\omega^{2},\omega^{3})\pmod{3}=(24,4)\otimes
(\omega^{2},e)\nonumber\\
\widehat{u}^{19}  &  =[\widehat{u}^{17},\widehat{u},\widehat{u}%
]=[(24,4)\otimes(\omega^{2},e),(4,9)\otimes(e,\omega),(4,9)\otimes
(e,\omega)]\nonumber\\
&  =(864,144)\pmod{25}\otimes(\omega^{3},\omega)\pmod{3}=(14,19)\otimes
(e,\omega)\nonumber\\
\widehat{u}^{21}  &  =[\widehat{u}^{19},\widehat{u},\widehat{u}%
]=[(14,19)\otimes(e,\omega),(4,9)\otimes(e,\omega),(4,9)\otimes(e,\omega
)]\nonumber\\
&  =(504,684)\pmod{25}\otimes(\omega,\omega^{2})\pmod{3}=(4,9)\otimes
(\omega,\omega^{2})\nonumber\\
\widehat{u}^{23}  &  =[\widehat{u}^{21},\widehat{u},\widehat{u}]=[(4,9)\otimes
(\omega,\omega^{2}),(4,9)\otimes(e,\omega),(4,9)\otimes(e,\omega)]\nonumber\\
&  =(144,324)\pmod{25}\otimes(\omega^{2},\omega^{3})\pmod{3}=(19,24)\otimes
(\omega^{2},e)\nonumber\\
\widehat{u}^{25}  &  =[\widehat{u}^{23},\widehat{u},\widehat{u}%
]=[(19,24)\otimes(\omega^{2},e),(4,9)\otimes(e,\omega),(4,9)\otimes
(e,\omega)]\nonumber\\
&  =(684,864)\pmod{25}\otimes(\omega^{3},\omega)\pmod{3}=(9,14)\otimes
(e,\omega)\nonumber
\end{align}

\begin{align}
\widehat{u}^{27}  &  =[\widehat{u}^{25},\widehat{u},\widehat{u}%
]=[(9,14)\otimes(e,\omega),(4,9)\otimes(e,\omega),(4,9)\otimes(e,\omega
)]\nonumber\\
&  =(324,504)\pmod{25}\otimes(\omega,\omega^{2})\pmod{3}=(24,4)\otimes
(\omega,\omega^{2})\nonumber\\
\widehat{u}^{29}  &  =[\widehat{u}^{27},\widehat{u},\widehat{u}%
]=[(24,4)\otimes(\omega,\omega^{2}),(4,9)\otimes(e,\omega),(4,9)\otimes
(e,\omega)]\nonumber\\
&  =(864,144)\pmod{25}\otimes(\omega^{2},\omega^{3})\pmod{3}=(14,19)\otimes
(\omega^{2},e)\nonumber\\
\widehat{u}^{31}  &  =[\widehat{u}^{29},\widehat{u},\widehat{u}%
]=[(14,19)\otimes(\omega^{2},e),(4,9)\otimes(e,\omega),(4,9)\otimes
(e,\omega)]\nonumber\\
&  =(504,6844)\pmod{25}\otimes(\omega^{3},\omega)\pmod{3}=(4,9)\otimes
(e,\omega)\nonumber
\end{align}

So we the relation $\widehat{u}^{31}=\widehat{u}$, such that the generator
$\widehat{u}$ possesses a strict cycle length of $30$. The ternary
multiplication of generators $\widehat{u}$ becomes%
\begin{equation}
\widehat{u}^{i}\widehat{u}^{j}\widehat{u}^{k}=\widehat{u}^{\left(
i+j+k\right)  \operatorname{mod}30},\ \ \ \ i,j,k\in\mathbb{N}_{0}%
,\ \ i,j,k\ \text{are odd}. \label{uu1}%
\end{equation}
Therefore, the ternary subgroup of the group field $\widehat{\mathrm{F}%
}^{\left[  \mathbf{6},\mathbf{3}\right]  }$ (\ref{f6}) generated by
$\widehat{u}=(4,9)\otimes(e,\omega)$ and unity $\widehat{\mathrm{e}}$
(\ref{ee}) is the cyclic ternary group $\mathsf{C}_{30}^{\left[  3\right]
}=\left\{  \widehat{u}^{i}\right\}  $ with multiplication satisfying
(\ref{uu1}), $\widehat{u}^{0}=\widehat{\mathrm{e}}$, and the unique
querelement $Q\left(  \widehat{u}^{i}\right)  =\widehat{u}^{\left(
30-i\right)  \operatorname{mod}30}$.

\textit{Recipient}'s public key $h$ is computed by the private key parameter
$\ell_{\mathtt{x}}=2$ as follows%
\begin{equation}
\widehat{h}=\widehat{u}^{2\ell_{\mathtt{x}}+1}=\widehat{u}^{5}=(9,14)\otimes
(\omega^{2},e).
\end{equation}

\textit{Sender} uses the key parameter $\ell_{\mathtt{k}}=3$ to compute the
ciphertext pair $(\widehat{c}_{1},\widehat{c}_{2})$ (taking into account that
$\widehat{u}^{35\operatorname{mod}30}=\widehat{u}^{5}$)%

\begin{align}
\widehat{c}_{1}  &  =\widehat{u}^{2\ell_{\mathtt{k}}+1}=\widehat{u}%
^{7}=(24,4)\otimes(e,\omega)\\
\widehat{c}_{2}  &  =[\widehat{\mathtt{m}},h^{2\ell_{\mathtt{k}}+1}%
,h^{2\ell_{\mathtt{k}}+1}]=[\widehat{\mathtt{m}},\widehat{u}^{35},\widehat
{u}^{35}]=[\widehat{\mathtt{m}},\widehat{u}^{5},\widehat{u}^{5}]\\
&  =(1134,1764)\operatorname{mod}25\otimes(\omega^{3},\omega^{4}%
)\widehat{\operatorname{mod}}3=(9,14)\otimes(e,\omega).
\end{align}

The finalized ciphertext pair $(\widehat{c}_{1},\widehat{c}_{2})=\left(
(24,4)\otimes(e,\omega),(9,14)\otimes(e,\omega)\right)  $ is sent to
\textit{Recipient} by \textit{Sender}. \textit{Recipient} initiates decryption
by computing from $\widehat{c}_{1}$ using the private parameter $\ell
_{\mathtt{x}}=2$.

Consider the $\ell_{\mathtt{k}}$-ternary power of the first ciphertext
component $\left(  \widehat{c}_{1}\right)  ^{2\ell_{\mathtt{x}}+1}$ and
compute its querelement having the general form $Q(\left(  \widehat{c}%
_{1}\right)  ^{2\ell_{\mathtt{x}}+1})=(a,b)\otimes(\omega^{c},\omega^{d})$.
Using the definition (\ref{q}), the relation yields%

\begin{align}
&  \lbrack(9,14)\otimes(\omega^{2},e),(9,14)\otimes(\omega^{2},e),(a,b)\otimes
(\omega^{c},\omega^{d})]\nonumber\\
&  =(126a,126b)\otimes(\omega^{c+2},\omega^{d+2})=(9,14)\otimes(\omega^{2},e)
\end{align}

Solving the modular equations for $a,b,c,d$ gives the unique querelement%
\begin{equation}
Q(\left(  \widehat{c}_{1}\right)  ^{2\ell_{\mathtt{x}}+1})=(9,14)\otimes
(e,\omega). \label{qq}%
\end{equation}

Finally, the original plaintext $m^{\prime}$ is recovered by performing
ternary multiplication on the ciphertext component $\widehat{c}_{2}$ with two
instances of the computed querelement (\ref{qq}) as in (\ref{c21})%

\begin{align}
\mathtt{m}^{\prime}  &  =\left[  \widehat{c}_{2},Q(\left(  \widehat{c}%
_{1}\right)  ^{2\ell_{\mathtt{x}}+1}),Q(\left(  \widehat{c}_{1}\right)
^{2\ell_{\mathtt{x}}+1})\right] \nonumber\\
&  =(1134,1764)\operatorname{mod}25\otimes(\omega,\omega^{2})\widehat
{\operatorname{mod}}3\nonumber\\
&  =(9,14)\otimes(\omega,\omega^{2})\equiv\mathtt{m}.
\end{align}

The numerical execution yields $\mathtt{m}^{\prime}=\mathtt{m}$, demonstrating
the validity of the proposed cryptographic framework within ternary group ring structures.

Now the dimension of any monomial element of the matrix ternary group field
$\widehat{\mathrm{F}}^{\left[  \mathbf{6},\mathbf{3}\right]  }$ is again four,
see (\ref{f6}). This gives minimal number of plaintext parameters, which can
be transferred. The secret message $\mathtt{m}$ can consist of the admissible
$N=5\ell_{N}+1$, where $\ell_{N}\in\mathbb{N}$ is number of composed $6$-ary
additions, in the binary case $N=\ell_{N}\in\mathbb{N}$, see (\ref{wln}). So
the number of formal summands in (\ref{f6}) is \textquotedblleft
quantized\textquotedblright, and the ternary cryptosystem algorithm can
transfer only the admissible $4\left(  5\ell_{N}+1\right)  $ plaintext
parameters at once.
\end{example}

\section*{Conclusions}

%\begin{quote}
Public-key cryptosystems eliminate the requirement for pre-shared secret keys
by enabling encryption with a publicly disclosed key and decryption with a
corresponding private key.

In this article we generalize the public-key cryptosystems to ternary
algebraic structures, with particular attention to ElGamal as representative
family.
%\end{quote}

The familiar ElGamal three-step pattern---publish a public key derived from a
private exponent, use ephemeral randomness to produce a two-part ciphertext,
and cancel the randomness at decryption---admits a direct analogue in ternary
algebraic settings. Replacing binary powers and inverses with $\ell$-ternary
powers and querelements yields a coherent encryption/decryption algebra in
which the decryption step recovers the plaintext by applying querelement
operations to the first ciphertext component and combining the result with the
second component.

Matrix ternarization provides a practical instantiation: mapping binary rings
and group rings to antidiagonal symbolic matrices produces nonderived ternary
unit groups closed under ternary multiplication while preserving additive
linearity. This construction increases per-element information density
(paired/plaintext vectors) and introduces additional algebraic structure that
an attacker must model, which can complicate straightforward adaptations of
binary discrete-log attacks.

Concrete numerical examples in the paper demonstrate correctness and
illustrate important structural features such as admissible word-length
quantization and cycle behaviour of ternary powers; these properties influence
message-space design and parameter selection. Remaining work includes
formalizing hardness assumptions for ternary analogues of discrete logarithm
and Diffie-Hellman problems, developing efficient and secure implementations
(including side-channel mitigations), and performing comparative performance
and cryptanalytic studies versus classical and post-quantum schemes.

Overall, the ternary framework is a viable algebraic generalization that
merits further cryptanalytic and implementation research before any practical deployment.
%\newpage
%\bibliographystyle{apsrev4-2}

\end{document}